\documentclass[structabstract]{aa}  
\usepackage{graphicx}
\usepackage{txfonts}
\usepackage{natbib}
\usepackage{longtable}
\newcommand{\kms}{\ensuremath{\mathrm{km\,s}^{-1}}}

\newcommand{\te}{\ensuremath{T_{\mathrm{e}}}}

\newcommand{\logl}{\ensuremath{\log(L/L_\odot)}}

\begin{document}

   \title{ Accurate age determinations of several nearby open clusters
     containing magnetic Ap stars}

   \author{J. Silaj
          \inst{1}
          \and
          J.D. Landstreet\inst{2,1}}

   \institute{Department of Physics and Astronomy, The University of
     Western Ontario, London, Ontario, N6A 3K7, Canada\\ 
     \email{jsilaj@uwo.ca} 
     \and
     Armagh Observatory, College Hill, Armagh, BT61 9DG, Northern Ireland     
     \email{jls@arm.ac.uk} 
           }

   \date{Received; accepted}

 \abstract
 { To study the time evolution of magnetic fields, chemical abundance
   peculiarities, and other characteristics of magnetic Ap and Bp
   stars during their main sequence lives, a sample of these stars in
   open clusters has been obtained, as such stars can be assumed to
   have the same ages as the clusters to which they belong. However,
   in exploring age determinations in the literature, we find a large
   dispersion among different age determinations, even for bright,
   nearby clusters. }
{ Our aim is to obtain ages that are as
 accurate as possible for the seven nearby open clusters $\alpha$~Per,
 Coma~Ber, IC~2602, NGC~2232, NGC~2451A, NGC~2516, and NGC~6475, each
 of which contains at least one magnetic Ap or Bp
 star. Simultaneously, we test the current calibrations of \te\ and
 luminosity for the Ap/Bp star members, and identify clearly blue
 stragglers in the clusters studied.}
 {We explore the possibility that isochrone fitting in the theoretical
   Hertzsprung-Russell diagram (i.e. $\logl$ vs. $\log \te$), rather
   than in the conventional colour-magnitude diagram, can provide more
   precise and accurate cluster ages, with well-defined uncertainties. }
 {Well-defined ages are found for all the clusters studied. For the
   nearby clusters studied, the derived ages are not very sensitive to
   the small uncertainties in distance, reddening, membership, metallicity, or
   choice of isochrones. Our age determinations are all within the
   range of previously determined values, but the associated
   uncertainties are considerably smaller than the spread in recent
   age determinations from the literature.  Furthermore, examination
   of proper motions and HR diagrams confirms that the Ap stars
   identified in these clusters are members,  and that the presently
   accepted temperature scale and bolometric corrections for Ap stars
   are approximately correct. We show that in these theoretical HR
   diagrams blue stragglers are particularly easy to identify.}
 {Constructing the theoretical HR diagram of a nearby open cluster
   makes possible an accurate age determination, with well defined
   uncertainty. This diagnostic of a cluster also provides a useful
   tool for studying unusual stars such as Ap stars and blue
   stragglers. }


   \maketitle
%

\section{Introduction}

Open clusters are created when giant molecular clouds undergo a burst
of star formation.  Despite differences in the length of time required
for stars of different masses to reach the ZAMS (zero-age main
sequence), the entire episode of star formation is relatively short
compared to the lifetime of a cluster, and hence stars in a single
open cluster are considered to have approximately the same age.  The
members of an open cluster are also kinematically linked; they share a
common space velocity (and therefore common proper motions and radial
velocities) which means that membership can be strongly tested for
stars that may be in nearby clusters.  Furthermore, it appears that
the stars of a given cluster have nearly identical initial chemical
compositions.  For these reasons, open clusters provide homogeneous
samples of stars that make demanding tests of stellar evolution models
possible, whereas such tests cannot be performed using samples of
field stars because of their generally uncertain distances, bulk
chemistry, and ages. Furthermore, the ages of stars on the upper main
sequence (since formation and arrival on the ZAMS) are much more
tightly constrained than is the case for field stars, even ones that
are rather close. Thus, cluster stars make excellent laboratories for
the study of evolution of stellar properties during the main sequence
phase.

The possibility of determining the ages of stars in open clusters with
reasonable accuracy has been exploited recently to study the evolution
of magnetic field strength in intermediate mass magnetic Ap and Bp
main sequence stars \citep{Bagetal06,lan07,Lanetal08}. By measuring
the strength of the magnetic fields in cluster Ap and Bp stars,
dividing the stars into mass bins, and comparing stars of a particular
mass bin but different ages (effectively using such a sequence of
stars of known ages as a proxy for the evolution of a single star of
that mass), it has been discovered that the magnetic fields of Ap and
Bp stars decrease strongly during the main sequence phase of
evolution.  Because age determinations of field Ap stars are simply
not accurate enough \citep{Bagetal06}, this remarkable discovery was
only made possible by instrumentation advances that made it practical
to detect and measure the magnetic fields of a reasonably large sample
of Ap and Bp stars that are members of open clusters. For this
project, where the ages of stars were obtained from those of a number
of clusters, both the precision and accuracy of the cluster ages was
critical. However, when we consulted the literature for cluster ages,
we found quite substantial disagreements, even for nearby open
clusters.  Our need for accurate and unambiguous ages led us to a
closer examination of possible methods for improving the determination
of open cluster ages.

Age determinations of open clusters have largely been based on the
fitting of theoretical isochrones to colour-magnitude diagrams
(usually using $M_V$ or $V$ vs. $B-V$) from optical photometry
(hereafter CMDs). Many such age determinations either do not provide
an associated uncertainty, or have stated uncertainties that are quite
large (often of the order of 0.3 dex, corresponding to an uncertainty
factor of about 2).  A number of different parameters can contribute
to the final uncertainty in the derived cluster age resulting from
such fitting, including uncertain distance modulus, uncertainties
about cluster membership and/or binarity of individual stars,
uncertain or variable reddening and extinction, uncertain cluster
average chemical composition (metallicity), and uncertainties from
fitting the isochrones to the available data. The complex interplay of
these parameters has stimulated interesting work on
maximum-probability methods of fitting cluster photometry of distant
clusters to isochrones \citep{NayJef06,Monetal10}; these methods not
only provide optimised ages but also provide clear guidance about
uncertainties of the various parameters.

In the case of young nearby clusters (within about 400~pc), however,
most of the parameters that enter the age determination using
isochrones are well determined. The distances to most nearby clusters
have been determined to within a few percent from Hipparcos
parallaxes of member stars, and thus the distance moduli are known
with an uncertainty of the order of 0.1~mag or better. Cluster membership,
determined from proper motions, and frequently parallaxes, for the
brighter stars, is much better constrained than in distant clusters
for which these data are not accurate enough to distinguish cluster
stars from field stars. In most nearby clusters, the reddening is small,
of the order of 0.1~mag or less, and is known within about 0.02~mag from the
intrinsic colours of hot cluster members. Many nearby young
clusters have reasonably well-known overall metallicity, and the
measured metallicities do not usually depart enough from solar values
to introduce age uncertainties even as large as 30\% (0.1~dex in
log(age)). Thus we could expect the ages of nearby young clusters
to be very accurately determined.

However, this is not the case. In Table~\ref{Tab_Cluster_Data}, we
list seven nearby open clusters that are of interest to us because of
the presence in each of one or more magnetic Ap or Bp star. In this
Table we also list some of the principal parameters of each cluster,
with our best estimate of the real uncertainties of each. In
Table~\ref{Tab_Cluster_Ages}, we have collected recent determinations
of the ages of these clusters.  In spite of general agreement on the
membership lists and the other main parameters affecting the cluster
ages, the median spread in age determinations is 0.4~dex, and in the
worst case is as large as 0.8~dex for IC~2602, so that the assigned
ages for this cluster range from about 10 million to more than 60
million years. Since many of the extreme ages (large and small) are
from relatively recent articles, in which there are no obvious errors
or use of clearly obsolete methods or data, we cannot simply choose
the best age determination for each cluster with confidence.

To understand this situation, we have tried to identify the aspects of
age determination for these optimal clusters which have introduced the
main variation from one work to another. 

One possible source of age scatter could originate in the variety of
computed evolution tracks and isochrones with which the observational
data are compared. The available isochrone sets
(e.g. \citet{Schetal92,Breetal93,Giretal00,Pieetal04,Breetal12}), and
consequently the ages deduced using them, differ from one another in
various significant ways. The tracks are based on different detailed
chemical compositions, even for nominally solar mixtures; the
opacities, equations of state, and nuclear reaction rates differ from
one grid of models to another, and overshooting from the convective
core is treated differently in different grids. The treatment of
atomic diffusion varies from one calculation to another, as do the
detailed numerical methods employed for the calculations. The result
is that even calculations for nominally identical chemistry and age
differ significantly. An example showing typical differences is found
in Figure~8 of \citet{Pieetal04}. Roughly speaking, the ages
associated with a given isochrone using different recent computations
made with similar composition and physics can differ from one another
by amounts of the order of 10\%, or about 0.05~dex. Furthermore, the
variation in the assumed metal mass fraction $Z$ appropriate to the
Sun by different groups (this can range from $Z_\odot \sim 0.0135$ to
$Z_\odot \sim 0.20$) contributes possibly as much as 20 or 25\%
uncertainty, or about 0.1~dex. The combination of these effects could
account for at least 0.1~dex variation in the age assigned to any
single cluster. However, this uncertainty, although significant, is not
nearly large enough to explain the observed scatter in assigned
cluster ages.

Since most of the common problems appear to contribute in only a minor
way to the uncertainties in the ages of these particular clusters, we
are forced to consider the actual isochrone fitting as a possible
major source of uncertainty. The published cluster ages are determined
mainly by comparison of the cluster member sequence in the $M_V$
vs. $B-V$ diagram to the best fitting isochrone. The best fitting
isochrone is selected primarily by comparison of the observations to
the so-called ``turnoff'', where the isochrone bends sharply away from
the main sequence, to describe the evolution up to and past the
terminal age main sequence (TAMS). In fact, we notice that this
comparison is very often ambiguous, especially in the case of young
clusters. This ambiguity is clearly illustrated in the left-hand panel
of Figure~\ref{fig:ic2602-2}, which is a CMD showing the extrema of
the ages found in the literature.  We emphasise that it is not a fit
to the data like the HRD depicted in the right-hand panel, but it is
provided to show 1) primarily that either of the two extreme
isochrones (and therefore the whole range in between, as well)
plausibly fit the data, and 2) that blue stragglers cannot be easily
distinguished from the normal members (which can add a significant
source of error in performing the age determination).  This figure
shows the very different ages that result depending on whether the
star to the left of the isochrone at about $M_V \approx -0.5$ is given
high weight, and whether the very bright star at about $M_V \approx
-3$ is considered a blue straggler or simply the most massive star in
the cluster. A similar ambiguity concerning which isochrone to select
is observed in the following two left panels, Figures~\ref{fig:2232}
and \ref{fig:2451A}. This ambiguity occurs primarily because
$B-V$ is not very sensitive to \te\ above about 10\,000~K.

The problem seems to be that the isochrones in the $M_V$ vs. $B-V$
diagram tend to be nearly vertical near the turnoff in young clusters,
and this feature contributes a quite significant uncertainty in the
choice of the best fit isochrone, depending on whether one chooses to
fit to an isochrone with a turnoff close to the most massive stars, or
one which fits the two or three hottest stars best but extends much
higher in the CMD. The combination of these factors leads to large
uncertainty in the age determination (usually given as $\log t$, where
$t$ is the time since formation, in years). 

In contrast to the CMD, the theoretical Hertzsprung-Russell diagram
(hereafter HRD) uses luminosity \logl\ vs. effective temperature log
\te.  In this description of the cluster HRD, both axes use stellar
 parameters that for bright clusters may usually be obtained from a
synthesis of multiple types of available information, for example
photometry in both the Str\"{o}mgren $uvby\beta$ system and the Geneva 
system.  Consequently, the values obtained for these parameters are
robustly determined, and the effects of peculiarities in the
spectrum of individual stars or variations in reddening across the
cluster may be handled using specialised temperature calibrations
and/or bolometric corrections and individual reddening corrections.

 A very useful feature of the theoretical HRD is that the turnoff of
 the isochrones is strongly hook-shaped at all ages. Consequently, if
 it is possible to identify on the upper main sequence the stars that
 are near the end of their main sequence evolution, the observed
 cluster main sequence is bracketed by only a narrow range of
 theoretical isochrones (see e.g. the RHS of
 Figure~\ref{fig:ic2602-2}).  As a result, working in the theoretical
 HRD often provides an unambiguous age for the cluster, and, equally
 importantly, a rather well-defined estimate of the uncertainty of the
 fit leading to that age.  It may be possible to achieve such
   results while still working in the CMD if different quantities are
   considered -- one of our referees suggests that isochrones in the
   $M_B$ vs. $U-B$ CMD would have similar properties
   to the isochrones in the HRD, and permit a similar unambiguous age
   determination for young clusters.
  
Our main objective in this paper, however, is to use the theoretical
HRD to try to determine ages for the clusters studied that are as
unambiguous and accurate as possible, and that have well-defined and
relatively small uncertainties. However, the study has also produced
some useful side products. First, we can confirm that the most recent
proper motions, parallaxes, and HR diagram positions of magnetic Ap
stars that are claimed to be cluster members do support cluster
membership.  Furthermore, the HR diagrams enable us to test current
calibrations of effective temperature and bolometric correction used
for the magnetic Ap stars, by comparing the positions of such stars to
those of normal middle main sequence stars (which presumably share the
same bulk structure and composition as the Ap stars) in cluster
HRDs. Finally, we find that the theoretical HR diagram is a very
suitable tool for identifying blue stragglers in young open clusters.

In the following section we discuss the sources of the data used in
this study. Section 3 considers the selection of candidate members,
and the determination of \te\ and \logl.  Section 4 outlines the
general process of determining cluster ages through the examination of
theoretical HR diagrams, while Section 5 discusses the ages we obtain
by employing this method for the seven individual clusters studied
here, and reports specific results concerning the Ap star members and
possible blue stragglers. A final section reviews the conclusions
reached in earlier sections.


\section{Data}
This paper investigates seven nearby open clusters: $\alpha$~Per,
Coma~Ber, IC~2602, NGC~2232, NGC~2451A, NGC~2516, and NGC~6475. The
clusters studied are listed in Table~\ref{Tab_Cluster_Data} and
Table~\ref{Tab_Cluster_Ages}.  

Table~\ref{Tab_Cluster_Data} provides a summary of basic geometric and
physical parameters of each cluster, including the distance $d$, the
mean parallax $\pi$, the true distance modulus $DM$, the mean colour
excess $E(B-V)$, the mean abundance $\log(N_{\rm Fe}/N_{\rm H}) - \log
(N_{\rm Fe, \odot}/N_{\rm H, \odot}) = [Fe/H]$, and the deduced heavy
element mass fraction $Z$. The references for most of these data are
provided below in the discussions of individual clusters. 

Cluster (logarithmic) ages derived from the literature are shown in
Table~\ref{Tab_Cluster_Ages}. Several recently published ages are
provided in the columns labelled $1 \dots 6$ and Others, with
references to the sources, almost all of which are relatively
recent. Our final derived ages, and their uncertainties, discussed
below, are in the final two columns. This table clearly illustrates
the problem of uncertainty of the ages of even these very nearby and
comparatively well-studied clusters: the derived ages range over a
factor of {\em more than 2} for $\alpha$~Per, IC~2602, NGC~2451A, and
NGC~2516.

All of the clusters share properties that should make accurate and
precise age determinations possible.  First, the open clusters studied
are all within 360~pc, and are therefore relatively bright.  Thus, the
brighter cluster members are fairly easy to observe, even with modest
instruments, and indeed many stars in each cluster are well studied
and have been observed in several photometric systems.  This allows us
to check the reliability and consistency of the measurements and the
derived fundamental stellar parameters.

The clusters in this study also all have securely determined
reddening; in most cases, they are known to have little or no
reddening and to lack differential reddening.  (It should be noted
that one cluster -- NGC~2516 -- does appear to have differential
reddening, but the effect is small and this cluster was particularly
desirable to include in the study because it contains several magnetic
Ap stars.)  

All of the clusters chosen for this study contain a fairly large
number of middle main sequence members in addition to their magnetic
star members.  This is useful for defining the observed main sequence
clearly, and it makes it possible to see how well the members fall
onto the theoretical isochrones, and with how much scatter.  Verifying
that the cluster is well-behaved in this manner helps to strengthen
our confidence that no large, systematic errors occur in our
determinations of \logl\ and \te.

As the clusters are all nearby, the individual stars in the selected
open clusters have their membership and distances well constrained by
astrometry.  The Hipparcos mission \citep{Esa97}, and its subsequent
extension to the Tycho-2 catalogue \citep{Hogetal00a,Hogetal00b}, have
been invaluable in providing high-precision proper motions and
parallaxes.  These datasets have allowed us to confirm many probable
members, and reject other candidates whose membership was dubious,
with much greater confidence than ever before.

The clusters chosen for this study contain stars for which photometric
data in the Str\"{o}mgren $uvby$ and/or Geneva photometric systems are
readily accessible from astronomy databases such as the SIMBAD
astronomical database (http://simbad.u-strasbg.fr/simbad/sim-fid), the
WEBDA database for open clusters
\citep[http://www.univie.ac.at/webda/,] []{MerPau03}, or the
University of Lausanne photometric catalog
\citep[http://obswww.unige.ch/gcpd/gcpd.html,] []{meretal97}.
Photometry in these two systems has been preferentially used because
of the availability of reliable temperature calibrations and
bolometric corrections for both normal and Ap stars.

\section{Method}

\subsection{Candidate Selection}

For each cluster, the initial selection of possible members was made
using three separate catalogs: \citet{rob99}, \citet{bau00}, and
\citet{dias01}.

The \citet{rob99} catalog is a census of cluster members of a number
of nearby clusters that is based on Hipparcos proper motions and
parallaxes. Typically, the dozen or so brightest cluster members are
identified. Members have been selected by an iterative process using
all available data, including parallaxes, proper motions, photometry,
and radial velocities. We have initially accepted all stars listed in
this article as cluster members.

\citet{bau00} have re-examined the Hipparcos catalogue to obtain mean
motions and distances of more than 300 clusters, and have used all
available data to establish membership of Hipparcos stars in these
clusters. This study provides a larger sample of clusters than
\citet{rob99}, but is limited to clusters more than about 200 pc away,
so only four of the seven clusters in this study are in that catalog.
However, the catalogue offers useful overlap and comparison with the
others. These authors provided membership probabilities based on all
available data which can be cross-referenced with other membership
probability calculations.  Initially all stars with non-zero
membership probability were accepted from this catalogue but again,
some clear non-members were later removed from the final HRDs.  A very
useful feature of the \citet{bau00} catalog is that known binary stars
are tagged.  When constructing an HRD, it is important to be aware of
binarity since (in the case of spectroscopic binaries or close visual
binaries treated as single stars by photometry) the increased
luminosity will shift the binary system's location upward above the
main-sequence relative to single stars by as much as 0.75~mag or
0.3~dex in \logl, possibly leading to an error
in the cluster age determination.

The catalog of open cluster members and cluster mean motions of
\citet{dias01} is based on Tycho-2 proper motion measurements, and
provides a much more extensive list of possible cluster members in the
clusters of interest to us than \citet{rob99}. Thus the proposed list
of members of each cluster spans a much wider range in temperature
and luminosity.  The number of fainter possible members is greatly
increased because of the fainter limiting magnitude of the Tycho-2
dataset compared to the Hipparcos data.  Membership probabilities for
this catalog are based essentially on proper motion data, and have
been determined using the statistical method of \citet{san71}. We note
that memberships based on Tycho~2 proper motions are somewhat less
secure than those based on Hipparcos motions because of the generally
somewhat larger uncertainties of the former. 

The clusters studied here are near enough that a substantial fraction
of middle main sequence members are Hipparcos stars, for which the
astrometric data on proper motions and parallaxes have been
substantially improved by \citet{van07}. The mean cluster parameters
and membership issues have been extensively rediscussed on the basis
of these new measurements by \citet{van09}. Unfortunately, this work
does not list the Hipparcos stars that are considered by van Leeuwen
to be members. However, when stars are selected from the revised
Hipparcos catalogue of \citet{van07} with very restrictive limits on
the range of parallax and proper motion, essentially the same lists of
stars are found to be members of the various clusters as result from
the studies above.

For assessing membership, it is essential to keep in mind that,
although the HIP stars of each cluster have astrometric data that are
tightly clustered in parallax and proper motion compared to nearby
field stars, the data for cluster members almost always show an
intrinsic scatter, particularly in proper motion, that is two or three
times larger than the uncertainties of individual astrometric data for
single stars would suggest \citep[see e.g. Figs. 10, 14, 16,
19, and 23 of][]{van09}. Confirming or rejecting membership on the
basis of astrometry thus requires comparison of data for individual
stars with the observed spread of data values for other cluster
members rather than simple measurement of the distance of a star's
data set from the mean cluster values in units of standard errors.

We have initially accepted as cluster members all stars with non-zero
membership probabilities in the three studies cited, but later removed
any stars that are clear outliers on the HRD and that have low ($<
20$\%) membership probabilities.  The remaining stars are thus all
probable cluster members. To further refine our member lists, the
astrometric data for the stars retained as cluster members have been
compared to the spread of astrometric data for HIP core members of
each cluster, and a few significant outliers have been removed.

The stars finally retained for this study, and shown in the various
figures, are listed in Table~\ref{Tab_Cluster_Members}. This table
lists the HIP number (if available) together with at least one other
name, the intrinsic colour $(B-V)_0$ and absolute visual magnitude
$M_V$, both corrected for the mean cluster reddening, and our adopted
values of \te\ and \logl, determined as described in the following
section.  Where one of the names provided is a cluster number
designation (e.g. NGC~6475~103), the number follows the convention of
of the WEBDA database. We note that these numbers are not always
correctly recognised by the SIMBAD database.

The entries for each cluster in this table appear somewhat
inhomogeneous. Only some of the stars in each cluster have cluster
number designations, and normally the $UBV$ photometry used for the
CMDs is not available for all stars. This is a consequence of the
relatively recent addition to cluster membership lists of stars in the
halos of the visible clusters which share the cluster proper motions
and parallax. Such stars can be added to membership lists with far
greater assurance with the data from the Hipparcos mission than was
possible before. These newly identified cluster members often have
Hipparcos numbers but lack cluster numbers.

\subsection{Determination of fundamental stellar parameters $\te$ and $\logl$}

Stellar effective temperatures were primarily determined from Geneva
photometry.  Geneva photometric values were given preference as they
represent a homogeneous set with good cluster coverage.  The six
Geneva colours $U, V, B1, B2, V1,$ and $G$ were obtained from the
University of Lausanne on-line photometric catalog and input into the
program CALIB, a fortran program that returns $\te$ and $\log g$
values based on the calibrations proposed in \citet{kun97} and
described there.  In this regime, the reddening of each star is
specified by the user for stars of cool and intermediate effective
temperatures (up to about 10\,500~K), in the case of hot stars,
the reddening is internally corrected for interstellar extinction.
Adopted reddening values were taken from the literature (generally, a
single value was adopted for the entire cluster membership, for
reasons outlined in the previous sections).  The reddening value for
each cluster is given in Table~\ref{Tab_Cluster_Data} and in its respective subsection of
Section~\ref{Results}.  Several previous investigators have
constructed colour-colour (e.g. $U-B$ vs. $B-V$) diagrams and
determined reddening by de-reddening the diagrams to line up with
locally calibrated main sequence stars (see e.g. \citet{dachkabu89}).
Since reddening is most often given in terms of $E(B-V)$, but the
Geneva program requires reddening in terms of B2 and V1, the
conversion $E(B2-V1) \simeq 0.88E(B-V)$ as used by \citet{hau93} and
given by \citet{gol80} was applied.

Str\"{o}mgren $uvby$ photometry values were also used to determine
$\te$ and $\log g$, and we opted to employ this method as a secondary
check on our values and to obtain values for a few clusters members
for which Geneva photometry did not exist.  The $uvby$ data
(specifically $V$, $b-y$, $m1$, $c1,$ and $\beta$) were taken from the
SIMBAD astronomical database and analysed by the program UVBYBETA of
Napiwotzki.  This program is based on the calibration of Moon and
others and is described in \citet{nap93}.  UVBYBETA uses an iterative
method to calculate the effective temperature ($\te$) and (log)
gravity of each star, and automatically calculates the individual
stellar reddening value $E(b-y)$. Recalling that $E(B2-V1) \simeq
1.15E(b-y)$ (see \citet{cra84}, \citet{luc80}, and \citet{cra76})
and using the previously given relation for $E(B2-V1)$ and $E(B-V)$,
$E(b-y) \simeq 0.765E(B-V)$.

Calculating the effective temperatures from two independent systems
provides a useful check.  In the majority of cases where both types of
photometry were available, the calculated effective temperatures
agreed within 2--300~K.  Since the average temperature of
our stars is around 10\,000~K, the dispersion between the two systems
is thus of the order of 3\%.  

As an additional check, a third source was consulted where possible.
\citet{mas06} use visual magnitude and 2-micron all-sky survey (2MASS)
infrared photometry to obtain effective temperatures of F, G, and
K-type stars (hereafter referred to as 2MASS temperatures), and claim
that their new method provides uncertainties on the order of 1\%.
While we consider this to be a very optimistic error estimate, the
values they obtained nevertheless provide a useful comparison.
Generally, very good agreement between the 2MASS temperatures and
$uvby$ and/or Geneva temperature determinations was found.

A minor difference between $uvby$ and Geneva photometry is
that the Geneva photometry is very homogeneous, and the tabular values
available from the Geneva photometry database are averaged values. In
contrast, there are often several slightly different individual
determinations of $uvby$, sometimes with outliers, and it was usually
not clear which values would be the best to use.

These considerations, in combination with the fact that for many stars
$uvby$ photometry was not available (this was especially true in
NGC~2232 and NGC~2451A), have led us to adopt the \te\ values derived
from Geneva photometry for all stars for which they were available,
with the other two photometric systems providing checks on the adopted
values, and \te\ values for the few stars without Geneva photometry.

To compute $\logl$ for the stars having $\te > 5500$~K, we used the
cluster distance modulus taken from \citet{van09} (with the exception
of NGC~2232, as discussed below) and individual visual magnitudes in
the standard distance-luminosity relation to obtain an absolute visual
magnitude $M_V$ for each star.  The effective temperature and absolute
magnitude were then used to compute the bolometric magnitude
(M$_{bol}$) according to \citet{bal94},

\begin{equation}
M_{bol}=M_V-5.5647+18.9446\theta-19.8227\theta^{2}+6.1302\theta^{3},
\end{equation}
where $M_V$ is the visual magnitude and 
\begin{displaymath}
\theta = \frac{5040}{\te}.
\end{displaymath}

For the few red giants in the clusters we obtained \te\ values from
\citet{Evansetal96}, \citet{Alonsoetal99}, and \citet{Villaetal09}. For
the red giants the bolometric corrections were taken from
\citet{Buzzetal10}.

Finally, the relation, \begin{equation} log \frac{L}{L_{\sun}} =
\frac{M_{bol\sun} - M_{bol}}{2.5} \end{equation} (where M$_{bol\sun}$
= 4.75) was used to calculate the desired luminosity parameter.  For
each cluster, the data sets of individual stellar fundamental
parameters are plotted together with theoretical evolution tracks for
stars of various masses and theoretical isochrones for $Z = 0.02$,
taken from \citet{Breetal93} and \citet{Beretal94}.

As one of the goals of this project has been to test the current
calibrations of \te\ and bolometric corrections for magnetic Ap stars
by placing these stars in the same theoretical HRDs as normal stars in
each cluster, the fundamental parameters of the apparent Ap cluster
members were obtained essentially as described by \citet{lan07}.
Briefly, \te\ values obtained from the same codes described above are
corrected as proposed by \citet{hau96} for Geneva photometry, and as
described by \citet{ste89} for Str\"{o}mgren photometry. Physically,
these corrections compensate approximately for the deficiency of UV
flux in the energy distributions of magnetic Ap stars compared to the
UV flux present in normal stars having the same Paschen continuum
slope, which leads to an overestimate of \te\ for a magnetic Ap star
when this is estimated using the Paschen continuum. Since the flux
contribution of the UV is small for late A stars and becomes more
important as \te\ rises, the corrections are small around 7000 or
8000~K and increase with \te. In contrast to our choice for normal
stars, when the (corrected) values of \te\ from the two types of
photometry were not very different, the values were averaged and
rounded.

Similarly, a slightly modified bolometric correction has been obtained
for magnetic Ap stars by \citet{lan07}, and this has been used for the
magnetic stars to derive the luminosities.

We have included in our CMD and HRD plots all cluster members from
Table~\ref{Tab_Cluster_Members} for which the necessary data were
available. However, because the photometric data available for various
cluster members is inhomogeneous, the CMD for a cluster may contain a
few stars not plotted in the HRD, and {\it vice versa}. Such stars
have blanks in appropriate columns in Table~\ref{Tab_Cluster_Members}.

\section{Age determination using theoretical HR Diagrams}

As discussed above, different treatments of the physics of the stellar
models that are used to derive theoretical isochrones and ages lead to
mildly different deduced ages associated with a particular
isochrone. This uncertainty cannot be resolved at present. Although
they incorporate the latest ideas and data, even the most recent
evolution models probably still are significantly different from the
structure of real stars. We estimate from comparisons of different
models that the current uncertainty in the absolute scale of ages
derived from models, for stars less than about $10^9$~yr old, is
probably still of the order of 0.1~dex in $\log t$. There may be
a small additional uncertainty associated with effects of rapid
rotation in some stars; on the other hand, as we shall see below,
possible magnetic fields seem to have little effect on the place of a
star in the HRD.

However, the relative age scale uncertainty between different clusters
fit to a single set of isochrones should be rather smaller than this
value, as the remaining error terms in the model computations probably
displace all evolution models of similar global structure (e.g. middle
main sequence stars) by similar amounts. This residual relative
uncertainty is probably of the order of 0.05~dex or less. Thus, it is not
particularly important which set of recent isochrones we use. We have
made use of several sets of evolution models and isochrones
described by \citep{Giretal00}, for metal mass fractions $Z = 0.008,
0.02,$ and 0.05, and our derived cluster ages are thus on the scale of
this data set.

Our fits do not include independent determination of reddening as
determined from individual B stars or colour-colour diagrams, or of
distance modulus as determined from Hipparcos parallaxes (except for
NGC~2232 as described below), because even without taking these as
free parameters we generally have very good fits of the isochrones to
the stars near the ZAMS in the HRD. Our essential single free
parameter is the cluster age. The clusters studied here have fairly
sparse upper main sequence membership, and so we are usually obliged
to focus on one or a few stars at the upper end of the HRD. Our
strategy is (1) to use the general shape of the isochrones near
the best fits to identify any high-mass blue stragglers; (2) to check
the most luminous stars for reports of binarity; (3) to pass
isochrones around the most luminous apparently normal and single star or stars,
determining the uncertainty of the fit using the isochrones that are
just a little too young or too old to intersect the error boxes of the
most luminous stars; and (4) to check the bracketing isochrones for
general consistency with the rest of the stars on the nearby main
sequence.

We typically find that this fitting procedure allows us to define a
unique cluster age with an uncertainty that is at most about
0.1~dex. This is, of course, an age which is based on the assumption
that the models underlying the isochrones have the same composition as
the cluster stars. In fact, the clusters under study appear to have
slight composition differences. To account for this, we use the
reported abundances of Fe to estimate the metal mass fraction $Z$ of
each cluster. Then, by comparison with a set of isochrones computed
for different $Z$ values, we approximately correct the ages determined
from a single set of isochrones.

All of the clusters studied here have had determinations of iron
abundance [Fe/H]. The reported values (discussed for individual
clusters below) range from about $-0.05$ to $+0.32$~dex, thus covering
a range of about a factor of two in iron abundance. We assume that, to
sufficient accuracy, the iron abundance is a proxy for overall value of
$Z$, and that the two quantities vary proportionally to one
another. According to \citet{Breetal12}, a reasonable choice of the
solar value of $Z_\odot$ is 0.0152, corresponding to an iron abundance
of $\log N_{\rm Fe}/N_{\rm H} = 7.52 - 12.00 = -4.48$. We then use the
reported iron abundances [Fe/H] to determine cluster $Z$ values by
adding or subtracting the (logarithmic) value of [Fe/H] to $\log
Z_\odot$.

From experiments we have carried out fitting a single cluster to
isochrones as discussed above, we find that when we fit a given
cluster, such as $\alpha$~Per, in the HRD with isochrones computed for
different values of $Z$, the deduced cluster age changes. Roughly, the
change in deduced age with respect to change in $Z$ relative to
isochrones with $Z$ value near 0.02 satisfies $d \log t/d \log Z
\approx -0.25$. Accordingly, we have determined a nominal age for each
of the clusters in our study by fitting with isochrones computed for
$Z = 0.02$, and then corrected this age for each cluster by an amount
deduced from the ratio of the cluster $Z$ to 0.02. The largest
log(age) corrections found are about 0.04, or ten percent, with an
uncertainty of less than 0.02 in general. Thus with the age
corrections, the cluster to cluster variations of composition
contribute only an unimportant amount to the total cluster age
uncertainty.

Our final age uncertainty is generally about $\pm 0.1$~dex. This
uncertainty is basically the uncertainty in the relative ages of the
clusters studied; it does not include the overall age scale
uncertainty due to mathematical and physical approximations made in
stellar modelling, which we have also estimated to be of the order of
0.1~dex. Thus, apart from the modestly uncertain overall scale of
isochrone based ages, we believe that we are able to determine cluster
ages to within about $\pm 30$\%. For the clusters studied here, this
result represents an important improvement in precision and accuracy
of the isochrone ages, especially compared to the current spread of a
factor of 2.5 (0.4~dex) in either direction (from 10 to 65 Myr) for
the age of IC~2602.

\section{HR Diagrams of individual clusters}
\label{Results}

\subsection{IC 2602 = Melotte 102} 

This is a young nearby stellar group embedded in the Sco-Cen OB
association, which is fairly distinctive on the sky, in proper motion,
and in parallax \citep{kha05}. It is at a distance of $d = 150.6 \pm 2.0$~pc
\citep{van09}, and has $E(B-V) = 0.04$ \citep{rob99,ran01}.  Previous
age determinations for this cluster vary widely, from $\log t = 7.0$
to 7.83, a factor of about 7 from lowest to highest
(Table~\ref{Tab_Cluster_Ages}). 

The assigned ages depend strongly on whether the brightest star,
$\theta$~Car = HD~93030, is considered as a member, and whether it is
considered a star with a normal evolutionary history or a blue
straggler.  $\theta$~Car is located in the central region of
IC~2602. The stellar parallax is consistent with the cluster parallax
at the $2 \sigma$ level, while the proper motions differ from mean
cluster proper motions by about 1~mas in each coordinate, typical of
the spread in proper motions of members of a nearby cluster
\citep[e.g.][Fig. 14]{van09}. The mean stellar and cluster radial
velocities, 20.2 and about 19~\kms, are very similar
\citep{van07,Hubetal08}. It thus appears very probable that
$\theta$~Car is actually a cluster member.  $\theta$~Car is a
chemically peculiar star, with C underabundant and N overabundant by
about 1~dex each with respect to solar abundances. It is also an SB1
system with a period of 2.2~d and a velocity semiamplitude of 19~\kms\
in which the secondary contributes negligibly to the total light
\citep{Lloetal95,Hubetal08}.

While the membership of $\theta$~Car seems probable, it seems unlikely
from our HRD that it is a normal member: there is too large a gap
between it and the next brightest member, the Be star p~Car =
HD~91465. An isochrone (of $\log t \approx 6.6$) passing through
$\theta$~Car misses the second brightest star by a large margin
and also falls significantly below the next five stars (see
Figure~\ref{fig:ic2602-2}).  If $\theta$~Car is a cluster member, then
it must have had a very abnormal evolutionary history. It has often
been argued that both the unusual place in the HR diagram, and the
chemical anomalies, are the result of mass transfer from the now
invisible secondary, so that $\theta$~Car is a blue straggler
\citep[e.g.][]{AhuLap07, Hubetal08}.  Thus we disregard $\theta$~Car
when we determine the age of IC~2602.

In this cluster the brightest normal member, HD~91465, which appears
somewhat evolved, is bracketed by $\log t = 7.4$ and $\log t = 7.6$
isochrones.  The cooler members all lie close to these two theoretical
isochrones.  Some of the B stars in the HR diagram scatter slightly
above the isochrones, but most of them are less than $2 \sigma$ above
them.  One or more stars might be unidentified doubles. Using the
constraint provided by the assumption that the second brightest
cluster member has undergone normal single star evolution, and so
should lie on an isochrone, we find the best fit log age of this
cluster to be $7.52 \pm 0.05$.

We next consider the various possible additional uncertainties of this
age. The reddening is only about 0.04~mag, and is uncertain by at most
about 0.02~mag. The distance is known from Hipparcos parallaxes with
an uncertainty of only about 3\% \citep{van09}, and so the true
distance modulus is uncertain by only about 0.03~mag. These
uncertainties are included in the uncertainties of the positions of
individual stars in the HRD; they mainly affect the vertical position
of the cluster main sequence. The general correctness of individual
star parameters and of their uncertainties is supported by the
excellent fit of the lower main sequence to the computed isochrone in
Figure~\ref{fig:ic2602-2}. These errors introduce no further
uncertainty in the age determination beyond the fitting uncertainty.

The main remaining uncertainty is the effect of chemical composition.
The iron abundance of IC~2602 is $-0.05 \pm 0.05$ \citep{Ranetal01},
corresponding according to our discussion above to $Z = 0.014$. This
leads to an increase in the deduced log(age) of $+0.04$~dex, which
changes the age by less than its fitting uncertainty, with an
uncertainty which is no more than half of the actual correction. We
conclude that the dominant uncertainty is still the fitting
uncertainty, and the age of IC~2602 (after correction for composition)
is $7.56 \pm 0.05$~dex or $36 \pm 4$~Myr. This age and its estimated error are
on the scale of the adopted isochrones, and the uncertainty is
primarily the uncertainty relative to ages of other clusters
determined on the same isochrone scale. If we include our roughly
estimated scale uncertainty due to approximations inherent in the
modelling process, the absolute age uncertainty is probably not much
larger than about $\pm 0.1$~dex, or $\pm 20$\%. 

In Table~\ref{Tab_Cluster_Ages} we list the ages determined here,
including the abundance corrections, and with uncertainties relative
to our adopted isochrone scale.

Two Ap stars are believed to be members of this cluster, HD~92385 (B9
Si) and HD~92664 (B9p Si). The proper motions and parallaxes of these
two stars are within the cloud of data points in $\mu - \pi$ space, so
we confirm the assumed cluster membership.  These two stars are both
directly detected to have magnetic fields
\citep{Bohetal93,Bagetal06,lan07}. A simple way to characterize the
field strengths of the magnetic stars is by using the RMS value of the
measurements of the mean longitudinal field $\langle B_z \rangle$. For
the two Bp stars the RMS fields are of the order of 400~G and 800~G,
respectively.  The rotation periods are, respectively, 0.55 and
1.67~d. Their positions in the HR diagram, derived as discussed above,
are consistent with the normal cluster stars and with the adopted
isochrone.

\subsection{NGC 2232}

This cluster, identified by a few bright stars and tight clustering in
proper motion space \citep{kha05}, is located at a distance of $d =
353 \pm 22$~pc, with $E(B-V) = 0.02$ \citep{van09,rob99}.  For this
cluster, almost no $uvby$ photometry was available, so the effective
temperatures computed from Geneva photometry could not be checked.
Using the cluster distance proposed by van Leeuwen, all the stars lie
above the theoretical isochrone. The simplest assumption is that the
cluster is actually somewhat nearer than the optimal value derived
from the re-reduced Hipparcos parallaxes. We have plotted the stars
assuming that the actual distance modulus is $1 \sigma$ smaller than
proposed by van Leeuwen, i.e. 7.59 rather than 7.73~mag. With this
choice the stars generally lie close to the isochrone.

The iron abundance of NGC~2232 is reported to be $0.27 \pm 0.08$~dex
above the solar value by \citet{MonPil10}, leading to an inferred
metallicity of $Z \approx 0.028$. We note that using isochrones
calculated for larger metallicity than the nominal value of $Z = 0.02$
would raise the position of the computed main sequence in the HRD, but
not by enough explain the discrepant position of the main sequence
with van Leeuwen's parallax value. We still need to adjust the
distance downward to fit the cluster main sequence.

Ages estimated for this cluster recently range from 7.35 to 7.73
(Table~\ref{Tab_Cluster_Ages}).  The brightest member, HD~45546 =
HIP~30772, is bracketed by $\log t = 7.5$ and $\log t = 7.6$
isochrones, leading to a nominal age, using the $Z = 0.02$ isochrones,
of $\log t$ for this cluster of $7.55 \pm 0.05$. Applying a
metallicity correction of $-0.04$~dex, our final log age is $7.51 \pm
0.05$, or about $32 \pm 4$~Myr.

No blue stragglers have been identified in this cluster
\citep{AhuLap07}, and none appear in our HRD. 

The Ap star in this cluster, HD~45583 (B9 Si), has astrometric data
which are fully consistent with membership in the cluster. The star
has a very large field, with an RMS value of about 2700~G, which
varies on a period of 1.177~d
\citep{Bagetal06,Kudetal06,lan07,Semetal08}. Its position in the HR
diagram is normal.

\subsection{NGC 2451A = Puppis Moving Group}

NGC~2451A is an indistinct nearby group of stars with well-defined
proper motion and parallax \citep{kha05,van09}. The cluster proper
motion region is well separated from the proper motion cloud of nearby
field stars, so membership is fairly unambiguous. The distance of this
cluster is $d = 185.5 \pm 3.7$~pc \citep{van09}, and the colour excess
is $E(B-V) = 0.01$ \citep{car99}.  The cluster photometry is generally
well-behaved, and good agreement between the $uvby$ and Geneva
temperatures was found wherever both kinds of photometry were
available.  Recent previous age determinations range from 7.3 to 7.78
(Table~\ref{Tab_Cluster_Ages}).

The iron abundance of this cluster is [Fe/H] = $0.02 \pm 0.08$
\citep{Hueetal04}, leading to $Z = 0.016$ and a log age correction
relative to $Z = 0.02$ isochrones of $+0.03$~dex. 

The brightest member of the cluster does not seem to be very evolved,
but is best fit by a $\log t = 7.6$ isochrone.  Several other less
evolved bright members may be better modelled by the $\log t = 7.8$
isochrone.  We note, however, that some of the brighter members of this
cluster are found to be binary stars, and hence may lie up to 0.75~mag
above the isochrone.  The nominal $\log t$ for this cluster is $7.70
\pm 0.07$, which we correct to $7.73 \pm 0.07$~dex, or about $54 \pm
8$~Myr.

According to \citet{AhuLap07}, it has been suggested that the stars
HD~61831 and HD~63465 are blue stragglers. The parallax and proper
motions of HD~61831 indicate that the star is, with high probability,
a member, but it lies on the isochrone of the cluster in our HRD, and
is not a blue straggler. The parallax and proper motions of HD~63465
are very different from those of the cluster \citep{van07}, and this
star is certainly not a member.

The one known Ap star member of the cluster, HD~63401, is an Ap Si
star. It is a HIP star, and its redetermined astrometric parameters
are fully consistent with cluster membership. It has a rotation period
of 2.41~d \citep{Henetal76}. This star has a firmly detected magnetic
field with an RMS value of about 350~G \citep{Bagetal06,lan07}. The
star occupies a normal place in the HR diagram.

\subsection{$\alpha$ Per = Melotte 20} 

This large, bright and nearby cluster occupies a very distinctive
place in proper motion and parallax space \citep{san01,kha05,van09}.
It is at a distance of $d = 177.6 \pm 2.7$~pc \citep{van09}, and has
significant reddening of $E(B-V) = 0.09 \pm 0.02$ \citep{rob99,san01}. The
cluster has a wide range of stars for which photometric measurements
are available, and using Geneva photometry to obtain \te\ values, most
members coincide very well with theoretical isochrones (see
Figure~\ref{fig:aPer}). Recent log age estimates range from 7.3 to
7.9, or from about 20 to 80 million yr (Table~\ref{Tab_Cluster_Ages}).

According to \citet{Boeetal03}, [Fe/H] = $+0.02 \pm 0.03$~dex for the
$\alpha$~Per cluster. This leads to an estimated cluster $Z = 0.016$
and a log age correction relative to $Z = 0.02$ isochrones of $+0.03$.

In the theoretical HRD, the age of this cluster is most strongly
constrained by the position of the brightest main sequence star in the
cluster, $\psi$~Per (HD~22192), which lies close to the TAMS.
\citet{van09} strongly argues that this star is indeed a member of the
cluster. If the bracketing isochrones are chosen to be just a little
younger and just a little older than values which just intersect the
error bars of this star, we find $\log t = 7.75 \pm 0.05$, which after
correction becomes $7.78 \pm 0.05$, or an age
of about $60 \pm 7$~Myr. The two bracketing isochrones also intersect
satisfactorily the error ovals of other stars near the turnoff (except
for two stars which may be unrecognised binary systems, and lie a bit
above the general trend), although these other stars do not constrain
the cluster age as tightly as $\psi$~Per.  The bright red giant after
which the cluster is named lies on the younger of the two bracketing
isochrones. This cluster is another example of a group whose age can
be strongly constrained by a single star near the TAMS.

According to \citet{AhuLap07}, the star $\delta$~Per = HD~92928 may be
a blue straggler. However, the proper motion $\mu_\delta$ is nearly 20
mas/a different from the motion of the cluster, so this star is
clearly not a cluster member. 

The one well-studied Ap star in $\alpha$~Per, HD~21699, is a B8 He
weak star with strong Mn. The new HIP astrometric parameters place the
star within the cloud of points of other cluster members. The star has an RMS
field of the order of 600~G and a rotation period of 2.49~d
\citep{Broetal85}. The position of HD~21699 in the HRD agrees well with
the neighbouring stars and with the theoretical isochrones.

The star HD~22401 has been suggested to be another Ap (SiSrCr) member
of the cluster. Its proper motion and parallax are consistent with
membership. However, unpublished observations made by one of the
authors with the ESPaDOnS spectropolarimeter at the
Canada-France-Hawaii Telescope indicate that this star has no
detectable magnetic field, and spectrum modelling shows that the star
has atmospheric abundances close to solar, so it is very probably not
a magnetic Ap star at all.

\subsection{NGC 2516}

NGC~2516 is a large, rich cluster with more than 300 members. Its
proper motion is similar to those of nearby field stars, but the
proper motions of members are confined to a small, well-defined cloud
in proper motion space. The cluster seems to be a richer version of
the Pleiades cluster, with a similar age.  NGC~2516 is at a distance
of $d = 345 \pm 11$~pc, and has $E(B-V) = 0.112 \pm 0.024$
\citep{sun02}.  The reddening of the cluster is mildly variable, as
characterized by the uncertainty given for the $E(B-V)$ value.
Differential reddening to this small degree affects temperature
determinations in a relatively minor way and should be more than
accounted for by the error bars associated with our \te\ estimates.
The large number of stars, and of Ap stars in particular, in this
cluster make it a very interesting agglomeration to study, despite the
variable reddening. Recent age determinations for this cluster show
less scatter than are found for most of the other clusters studied
here, and range from 7.8 to 8.2 (Table~\ref{Tab_Cluster_Ages}),
probably because the three red giant members of the cluster can be
used to strongly constrain the appropriate isochrone in the CMD.

\citet{Maietal11}  report that [Fe/H] = $+0.01 \pm 0.07$, leading to
$Z = 0.016$ for this cluster. The corresponding correction to ages
derived from $Z = 0.02$ isochrones is $+0.03$. 

Although the $uvby$ and Geneva temperature determinations were in good
agreement for this cluster, there is an obvious scatter of the stellar
data about the isochrone. It is not completely clear why this cluster
exhibits what seems to be ``excess'' scatter; this scatter is strong
above $\te = 10\,000$~K, where correction for reddening is built into
both temperature calibrations.  Possibly the cluster has an abnormally
large number of unrecognized binary systems.  In addition, the
membership list is less secure than for most of the other clusters
studied here; as mentioned above, the proper motions of cluster
members are very similar to the motions of the field stars in this
region of the sky \citep{kha05}, and the cluster is distant enough
that measured parallaxes do not provide a strong discriminant against
foreground or background non-members.

The majority of stars in the middle of the diagram would be best fit by
the lower isochrone at $\log t = 7.9$, but the brighter members near
the turnoff clearly adhere more closely to the upper bracketing
isochrone of $\log t = 8.1$ (see Figure~\ref{fig:2516}).  We
adopt a corrected age of $\log t = 8.08 \pm 0.1$, about $120 \pm 25$~Myr. This
age is supported by the fit of the cluster's three red giants to the
isochrones. 

This age estimate disregards the star HD~66194 = HIP~38994 = V374~Car,
seen in the upper left corner of Figure~\ref{fig:2516}. This star has
been identified as a blue straggler in the list of
\citet{AhuLap07}. The proper motions and parallax of the star from the
new reductions of the Hipparcos data \citep{van07} place it well
within the cloud of proper motion points for cluster members shown in
Figure~19 of \citet{van09}, and the stellar parallax is consistent
(within less than $1 \sigma$) with the cluster distance. HD~66194 is
apparently a spectroscopic binary \citep{GonLap00}, but since the
period is not known, we cannot use the stellar radial velocity to test
membership. However, in the HRD the position of HD~66194 is consistent
with membership, assuming that this star has had abnormal evolution.
We conclude that on balance it is very likely that this star is indeed both
a genuine cluster member and a blue straggler.

The cluster contains several stars which have been classified as
magnetic Aps. None of these stars are Hipparcos stars, and so none
have newly revised, high-accuracy proper motions. Instead, the proper
motions are obtained from the Tycho data. The astrometric properties
of all these Ap stars are consistent with cluster membership, but
since the uncertainties of the measured motions are of the order of 3
or 4~mas/a, membership is not nearly as secure as for the majority of
Ap and Bp stars studied in this work. 

Magnetic fields have been detected in four of the cluster Ap and Bp
stars \citep{Bagetal06,lan07}. HD~66318 (A0p SrCrEu), the coolest of
these four, has a huge RMS field of about 4400~G, and has been studied
in detail by \citet{Bagetal03}. The three hotter Ap stars with
detected fields, HD~65712 (A0p Si 4200), HD~65987 (B9p Si), and
HD~66295 (B8/9p Si), all have RMS fields in the range of
500--900~G. Rotation periods have not been established for any of
these four stars with real certainty, but that of HD~66295 is probably
2.45~d. The remaining Ap stars in NGC~2516 plotted in
Figure~\ref{fig:2516} have been surveyed for fields once or twice
each, but the fields were below the threshold for detection.  The
magnetic Ap stars scatter about the bracketing isochrones about as
well, or as poorly, as the normal stars.

\subsection{NGC 6475 = Messier 7}

NGC~6475 is a sufficiently bright and obvious cluster to have
attracted the attention of Charles Messier in the 18th century. It is
at a distance of $d = 302 \pm 10$~pc \citep{van09}, and has $E(B-V) =
0.06$ \citep{mey93,rob99}.  It should be noted that membership
determinations for the cluster are hampered by the fact that the mean
motions of the cluster are very close to the mean motions of field
stars in the same direction, although the cluster motions are tightly
bunched.  Perhaps for this reason, several stars that have been
thought to be probable members in previous studies appear from their
positions in our HRD or from discrepant proper motions to be
non-members (for example, HD~162205 = NGC~6475~3 lies far enough below
the isochrones of Figure \ref{fig:6475} to be regarded as a probable
non-member).

The metallicity, according to \citet{Sesetal03} is $+0.14 \pm 0.06$,
slightly above solar. The corresponding $Z = 0.021$, and the age
correction to isochrone ages from $Z = 0.020$ computations is
$-0.01$~dex. 

It was found that the temperatures derived from the calibrations of
Geneva $XY$ and $pTpG$ photometry were somewhat discrepant for a few
stars of this cluster that have \te\ values right at the boundary
between the two calibrations, at about 10\,000~K. In such cases an
average of the \te\ values produced by the two calibrations was
generally taken. Two stars that lie above and to the right of the
isochrones, marked by asterisks, are known to be spectroscopic
binaries, and a third star, HD~162781, that is separated from other
stars inside the curve of the isochrone, may also be a binary. 

Previous age determinations (Table~\ref{Tab_Cluster_Ages}) range from
$\log t = 8.22$ to 8.55. Our best (corrected) age estimate, considering the
breadth of the main sequence in our HRD, is $8.34 \pm 0.10$, or about
$220 \pm 50$~Myr.  This cluster is old enough that age determinations
in the CMD benefit from the more hooked shape of isochrones (compared
to those for young clusters) as stars evolve towards the TAMS, so that
working in the theoretical HRD does not increase the precision of age
estimation as much as for younger clusters. In addition, red giant
members help to define the age in both the CMD and the HRD. 

According to the summary of \citet{AhuLap07}, two stars have
previously been identified as blue stragglers: HD~162374 and
HD~162586. Both stars have Hipparcos parallaxes and motions that are
consistent with membership \citep{van07}. The brighter of these stars,
HD~162374, is particularly interesting. If it is indeed a member, its
location in our HR diagram certainly identifies it as a blue
straggler, similar to $\theta$~Car in IC~2602, and to HD~66194 in
NGC~2516. HD~162586 may be a different kind of blue straggler. Its
position in our HRD is significantly hotter than the stars that
cluster about the isochrone, and near the TAMS. This star may be a
mild blue straggler that is completing its delayed main sequence
evolution.

Our HRD shows two other stars that appear to be blue
stragglers. HD~161649 and HD~163251 both appear to be fairly secure
members, but both are substantially too hot to lie on the cluster
isochrones. Both appear to be still close to the ZAMS. 

This cluster contains several Ap stars, among the largest number found
in any of the clusters in this study. The chemical abundances of
several of these stars have been studied by \citet{Foletal07}.  Four
Aps are plotted in Figure~\ref{fig:6475}. HD~162576 (B9.5p SiCr) and
HD~162725 (A0p SiCr) are both HIP stars, with revised proper motions
and parallaxes that are consistent with membership.  Both have RMS
fields of a few tens of G, firmly detected with ESPaDOnS at the
CFHT. The fields of HD~162305 (B9p) and HD~162588 (Ap Cr) are so far
undetected \citep{Lanetal08}. The Tycho motions of these two stars are
consistent with membership; since these particular Tycho data have
error bars only a little larger than 1~mas/a, this is a fairly strong
result. The small field values of Ap stars in this cluster, compared
to those mentioned above for stars in younger clusters, are consistent
with the general decline in field strength with age found by
\citet{Lanetal08}. These stars all lie within in the general spread of
normal stars about the isochrones.

\subsection{Coma Ber = Melotte 111} 

The cluster in Coma Berenices is a sparse group of stars at high
galactic latitude. It is at a distance of $d = 86.7 \pm 0.9$~pc
\citep{van09}, with a colour excess of $E(B-V) = 0.00$
\citep{rob99}. Although the cluster mean motions lie well within the
cloud of field star proper motions in this direction, the cluster has
a very tight distribution compared to the field, and membership can be
strongly tested. 

\citet{Gebetal08} find that the Fe abundance in this cluster is
essentially solar, so the value of $Z = 0.015$ and the age correction
is about $+0.03$~dex. 

The cluster is the oldest cluster we investigated, and thus the main
sequence has only stars later than about A2 (Figure~\ref{fig:cber}).
We found that the $uvby$, Geneva, and 2MASS temperatures were in
good agreement with one another and placed the stars on the theoretical
isochrone.  In this cluster, the final plot includes only stars for
which Geneva photometry is available. The range of ages proposed in
recent work goes from 8.65 to 8.9; as in NGC~6475, the advanced age of
this cluster means that in the photometric CMD the isochrones have a
sharp hook near the top of the main sequence and so ages are
relatively precisely determined directly from CMDs. From the best fit
isochrones in the HRD, our corrected adopted age is $\log t = 8.75 \pm
0.07$, or about $560 \pm 90$~Myr.

\citet{AhuLap07} report that the star HD~108662, a magnetic Ap star,
has been identified as a blue straggler. However, the proper motions
of this star, as reported by \citet{van07}, are completely inconsistent
with membership. This star is not a member of Mel~111. 

The one well-established magnetic Ap star in the cluster is HD~108945
= 21~Com, an A2p SrCr star with a rotation period of 2.0 days and a
weak but unambiguously detected magnetic field with an RMS value of
around 150~G \citep{Lanetal08}. The revised astrometric parameters of
this star confirm that it is a cluster member. This star is currently
one of the most evolved main sequence stars in the cluster. Its
position on the isochrone is normal.

{\section{Summary and Conclusions}

This work started from the need for precise and accurate age
determinations for a number of nearby clusters in order to obtain
accurate relative elapsed ages since the ZAMS of magnetic stars in
these clusters. We observed that recent measurements of the ages of
these clusters, based on stellar evolution calculations and isochrone
fitting,  can differ by factors of 2 or 3, even though membership in
the clusters is largely well-established, the distances are accurately
known, and there is little reddening. 

We have tried to determine the sources of the wide range of ages
published for such clusters as $\alpha$~Per and IC~2602. We considered
the possible effects of differences in cluster membership and cluster
parameters, as well as differences between various isochrone
computations and those due to differences in chemical compositions
between clusters. 

We were unable to find any reasonable source for the spread in
reported ages except for the various ways in which the isochrones are
fitted (in the $M_V$ or $V$ vs. $B-V$) colour-magnitude diagram,
or CMD). We have experimented with isochrone fitting in the
theoretical Hertzsprung-Russell diagram (HRD) and we conclude that
fitting cluster photometry in the HRD provides an unambiguous
way to determine the ages of these particularly favourable clusters,
and the uncertainties in these ages. This is because the isochrones
for different ages, especially for young clusters, do not diverge
sharply in the CMD from one another, but mostly extend to differing
upper absolute magnitudes, so that isochrones of quite different ages
can plausibly fit the same data. In contrast, for all ages the
isochrones in the HRD have a fairly sharp bend at the high end of the
main sequence which makes fitting to stellar data less ambiguous, as
long as there are one or two stars that are at least about 50\% of the
way through their main sequence evolution.

Furthermore, the behaviour of the isochrones in the HRD is such that
in most cases a reasonably clear assignment of the fitting
uncertainty is easily possible, and this uncertainty is often
remarkably small, of the order of $30$\%  or less (0.1~dex). 

The advantage of working in the HRD gradually diminishes with cluster
age, as the isochrones in the CMD become more sharply bent, and is not
great for cluster ages above perhaps 200~Myr. 

We have re-determined the isochrone ages of seven nearby clusters. We
conclude that our age determinations are precise (using a single set
of isochrone computations, but including the effects of small
differences in cluster compositions) to within less than about $\pm
0.1$~dex, or about 30~\%, with a possible additional overall scale
uncertainty (that does not affect the relative age determinations) due
to uncertainties in stellar modelling physics, also of about
0.1~dex. These results represent a significant improvement in the
determination of ages of most of the clusters studied.

A subsidiary aim of this work has been to test the current calibrations
for obtaining \te\ values of magnetic Ap stars from photometry. In
general, the Ap stars in the clusters studied lie well within the
spread of normal stars about the isochrones in the clusters
studied. This result supports the adopted temperature calibration for
these stars, although only at roughly the 10\% level in the
temperature range between about 9500~K and 15000~K. The HRD positions we
find for the Ap stars provide useful tests of cluster membership, and
as discussed above are generally consistent with the claim that these
stars are cluster members.

We find that our method of age determination facilitates the
identification of blue stragglers in such favourable clusters, and we
confirm the blue straggler nature of stars in IC~2602, NGC~2516, and
NGC~6475, and show that several other proposed blue stragglers are not
in fact cluster members. 

Our overall conclusion is that where enough information is available
to permit the secure determination of cluster membership, reddening,
and chemical composition, and the stellar fundamental parameters \te\
and \logl, studying clusters in the theoretical (\te\ -- \logl)
Herzsprung-Russell diagram can provide valuable new insights, and
securely determined stellar and cluster ages.  Our methods can
usefully be applied to obtain accurate ages for at least a dozen other
nearby clusters (closer than, say, 400~pc) that contain several
Hipparcos stars and for which accurate parallaxes, mean proper
motions, radial velocities, multi-colour photometry, and reddening are
available \citep{rob99,bau00}.

\clearpage
\begin{table*}[h]
\caption[]{\label{Tab_Cluster_Data} Basic data for clusters studied.}
\begin{tabular}{lcccccc}
\hline\hline
Cluster & parallax & DM & distance (pc) & $E(B-V)$ & [Fe/H] & $Z$ \\
\hline
$\alpha$ Per & $5.63 \pm 0.09$ & $6.18 \pm 0.03$ & 177.6 & $0.09 \pm 0.02$ & $+0.02 \pm 0.03$ & 0.016 \\
Coma Ber & $11.53 \pm 0.12$ & $4.69 \pm 0.02$ & 86.7 & $0.00 \pm 0.02$ & $0.00 \pm 0.07$ & 0.015 \\
IC 2602 & $6.64 \pm 0.09$ & $5.86 \pm 0.03$ & 150.6 & $0.04 \pm 0.02$ & $-0.05 \pm 0.05$ & 0.014 \\
NGC 2232 & $2.83 \pm 0.17$ & $7.59 \pm 0.14$ & 353 & $0.02 \pm 0.02$ & $0.27 \pm 0.08$ & 0.028 \\
NGC 2451A & $5.39 \pm 0.11$ & $6.32 \pm 0.04$ & 185.5 & $0.01 \pm 0.02$ & $0.02 \pm 0.08$ & 0.016 \\
NGC 2516 & $2.90 \pm 0.08$ & $7.68 \pm 0.07$ & 345 & $0.11 \pm 0.03$ & $0.01 \pm 0.07$ & 0.016 \\ 
NGC 6475 & $3.31 \pm 0.13$ & $7.16 \pm 0.08$ & 302 & $0.06 \pm 0.03$ & $0.14 \pm 0.06$ & 0.021 \\
\hline\hline
\end{tabular}
\end{table*}

\begin{table*}[h]
\caption[]{\label{Tab_Cluster_Ages} Comparison of cluster ages ($\log t$).}
\begin{tabular} {lccccccccc}
\hline \hline
Cluster    & 
1          & 2 & 
3          & 4 & 
5          & 6 & 
Other(s)   & $\log t$  & $\sigma$
\\ \hline
$\alpha$ Per & 7.55 & 7.854 & 7.71 & 7.72 & 7.3 &  & 7.72$^7$,
   7.7$^8$, 7.89$^9$, 7.81$^{10}$ & 7.78 & 0.05\\ 
Coma Ber & 8.78 & 8.652 &  &  &  &  & 8.9$^8$ & 8.75 & 0.07\\ 
IC 2602 & 7.83 & 7.507 & 7.56 &  & 7.0 & 7.50 & 7.48$^{11}$ & 7.56
   & 0.05\\ 
NGC 2232 & 7.49 & 7.727 & 7.35 &  &  &  & 7.51$^{12}$ & 7.51 & 0.05\\ 
NGC 2451A & 7.76 & 7.780 & 7.56 &  & 7.3 &  & 7.78$^{13}$, 7.70$^{14}$ & 7.73 & 0.07\\ 
NGC 2516 & 8.08 & 8.052 & 8.03 & 8.15 &  & 7.80 & 8.10$^{12}$, 8.20$^{15}$,
   8.20$^{16}$ & 8.08 & 0.10\\ 
NGC 6475 & 8.22 & 8.475 & 8.35 & 8.35 &  & 8.55 & 8.34$^{17}$, 8.30$^{18}$ & 8.34 & 0.10\\ 
\hline\hline
\end{tabular}
\\
\tablebib{(1)~\citet{kha05};  
(2) \citet{lok01};           (3) \citet{mer81};           
(4) \citet{mey93};
(5) \citet{san01};           (6) \citet{tad02};
(7) \citet{mak06};           (8) \citet{pin98};
(9) \citet{mar01};           (10) \citet{bas99};
(11) \citet{sta97};          (12) \citet{lyr06};
(13) \citet{pla01};          (14) \citet{car99};
(15) \citet{sun02};          (16) \citet{bon05};
(17) \citet{Sesetal03};      (18) \citet{Villaetal09}.
}
\end{table*}
\clearpage

\begin{figure*}
\centering
\includegraphics*[width=3.45in]{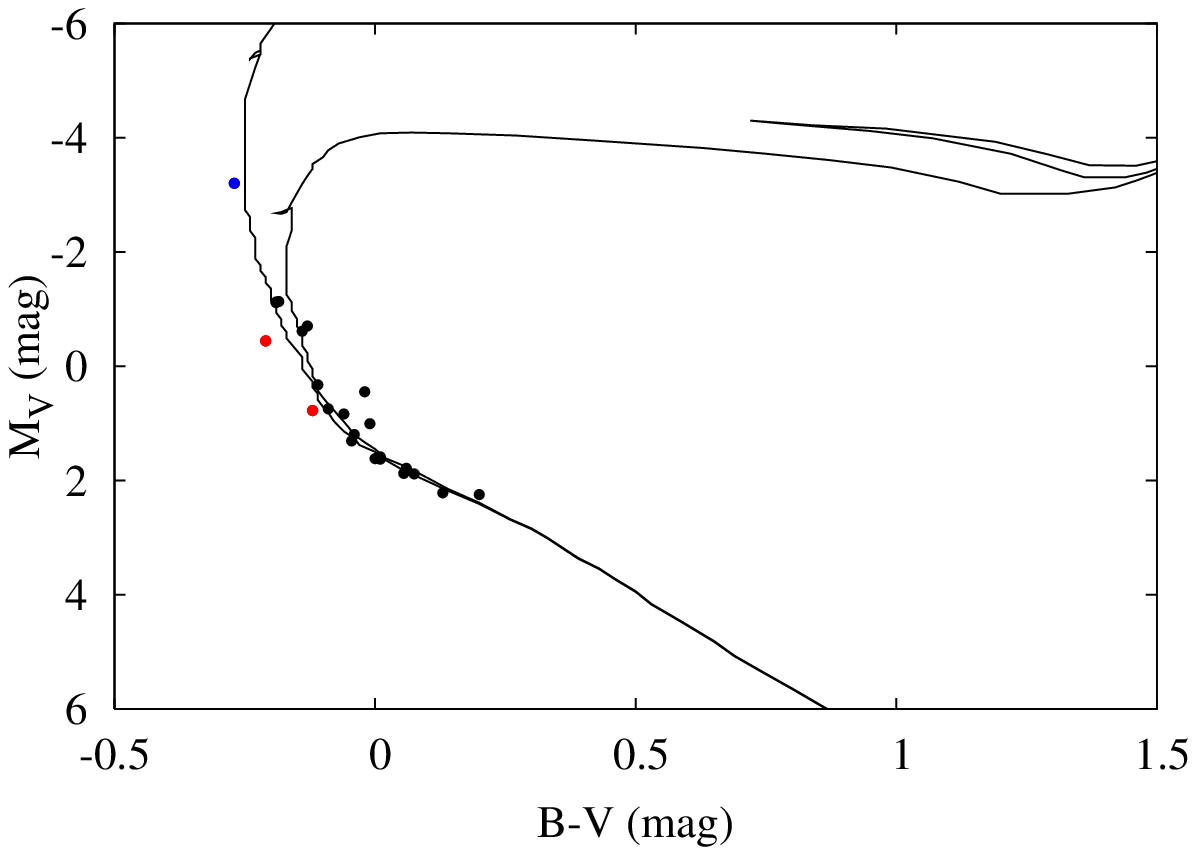}
\includegraphics*[width=3.45in]{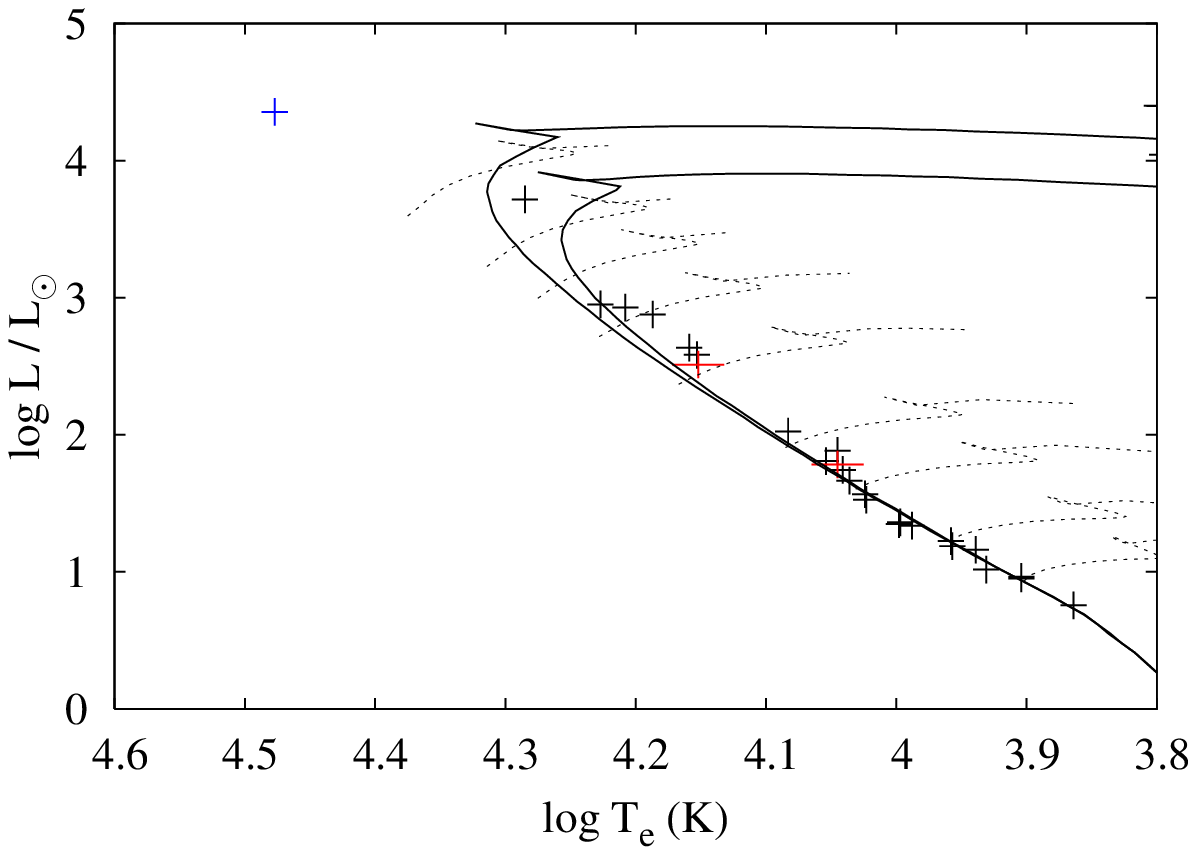}
\caption{Left Panel: Colour-magnitude diagram for IC~2602.
    Probable (normal) cluster members are plotted with filled black
    circles while Ap star members and a blue straggler are represented
    by filled red and blue circles, respectively. Bracketing
  isochrones corresponding to the limiting values found in the literature from Table \ref{Tab_Cluster_Ages} ($\log t = 7.0$ (leftmost) and $\log t =
  7.8$ (rightmost)) are represented by solid lines.  Right Panel:
  Theoretical HRD for IC~2602. Normal cluster members are plotted
    in black as points with error bars, Ap stars are again
    distinguished by red, and the blue straggler by blue, as in the
    left panel. Bracketing isochrones corresponding to $\log t = 7.4$
  (leftmost) and $\log t = 7.6$ (rightmost) are represented by solid
  lines.  Evolution tracks (from bottom right and progressing to upper
  left corner) for $1.7, 2.0, 2.5, 3.0, 4.0, 5.0, 6.0, 7.0,$ and
  $9.0M_{\sun}$ stars are represented by dotted lines. The apparent
  blue straggler is HD~93030 = $\theta$~Car.}
\label{fig:ic2602-2}
\end{figure*}

\begin{figure*}
\centering
\includegraphics*[width=3.45in]{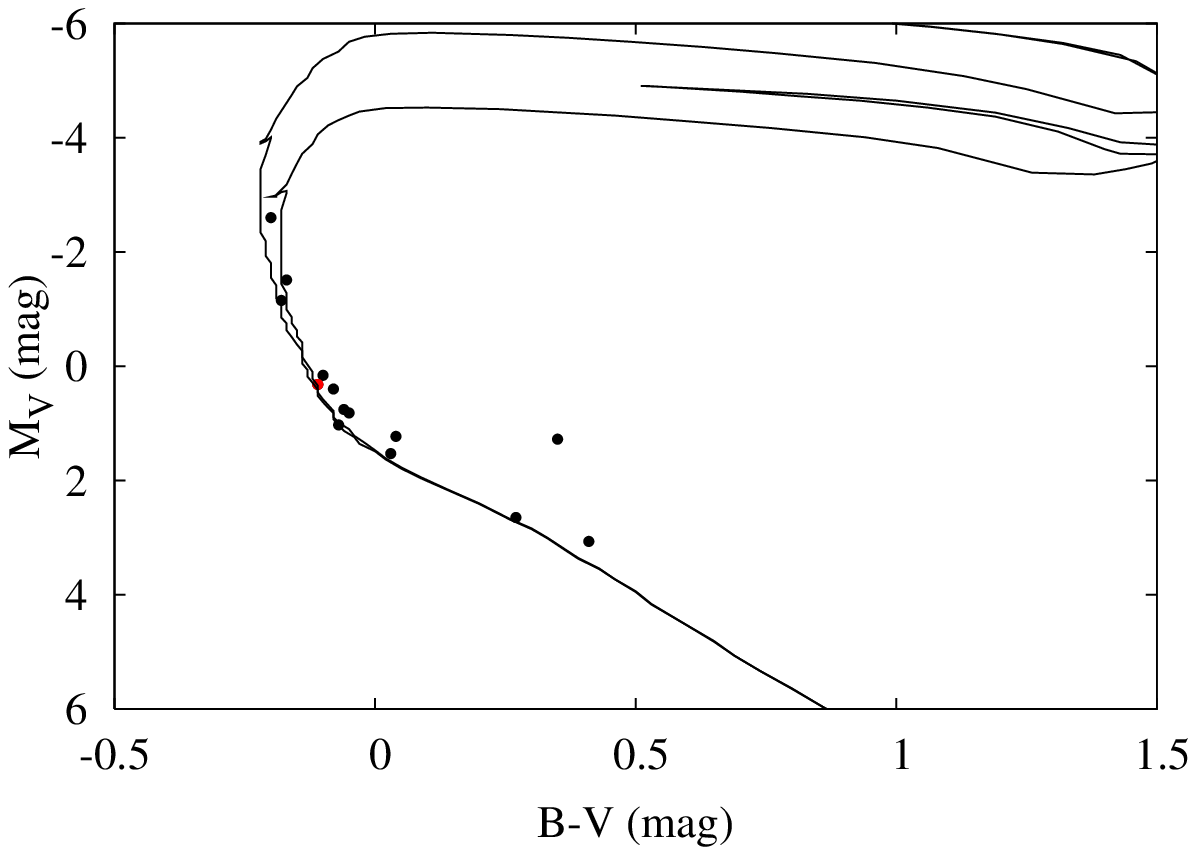}
\includegraphics*[width=3.45in]{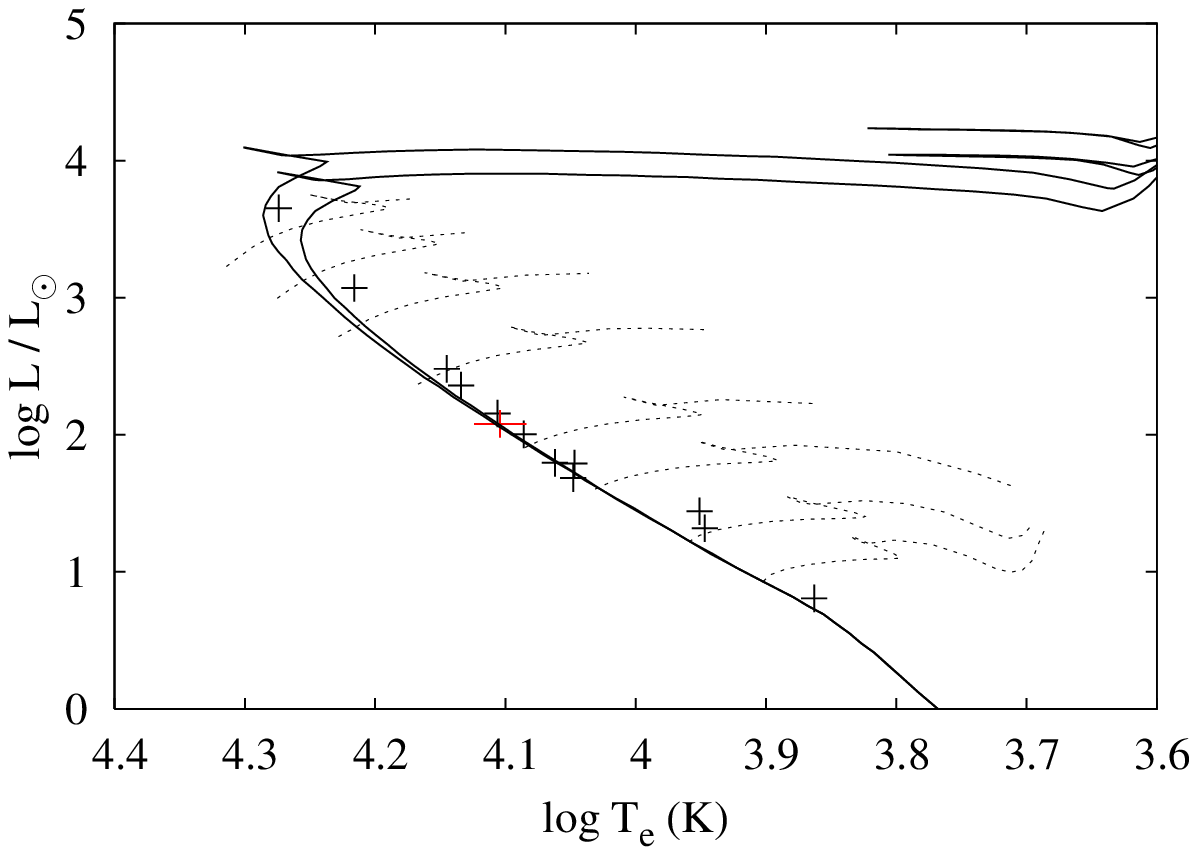}
\caption{Left Panel: Colour-magnitude diagram for NGC~2232. Symbols and colours
  are as in Figure~\ref{fig:ic2602-2} with the bracketing isochrones
  having values of $\log t = 7.4$ (leftmost) and $\log t = 7.7$
  (rightmost). Right Panel: Theoretical HRD for NGC~2232. Symbols are
  as in Figure~\ref{fig:ic2602-2}. Bracketing isochrones corresponding
  to $\log t = 7.5$ (leftmost) and $\log t = 7.6$ (rightmost) are
  shown.  Evolution tracks (from bottom right and progressing to upper
  left corner) for $1.7, 2.0, 2.5, 3.0, 4.0, 5.0, 6.0,$ and $7.0M_{\sun}$
  stars are shown.}
\label{fig:2232}
\end{figure*}

\begin{figure*}
\centering
\includegraphics*[width=3.45in]{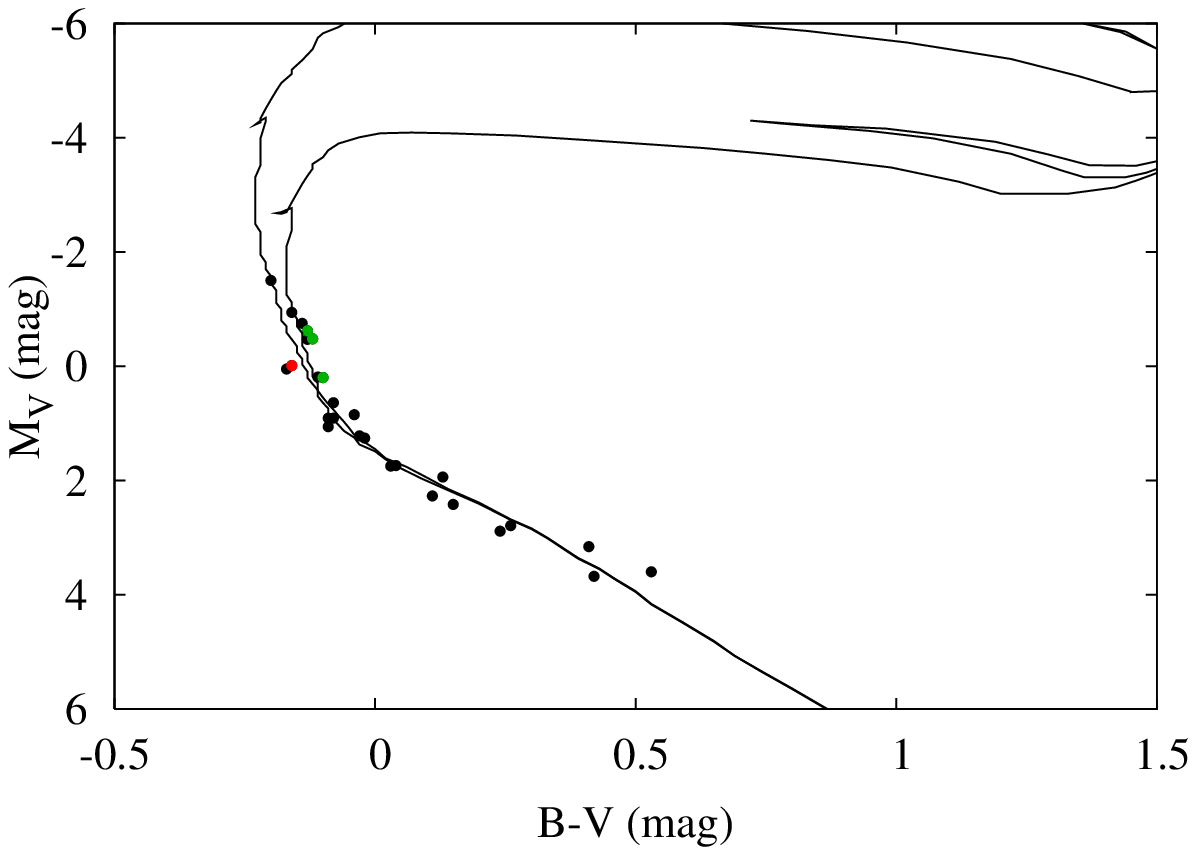}
\includegraphics*[width=3.45in]{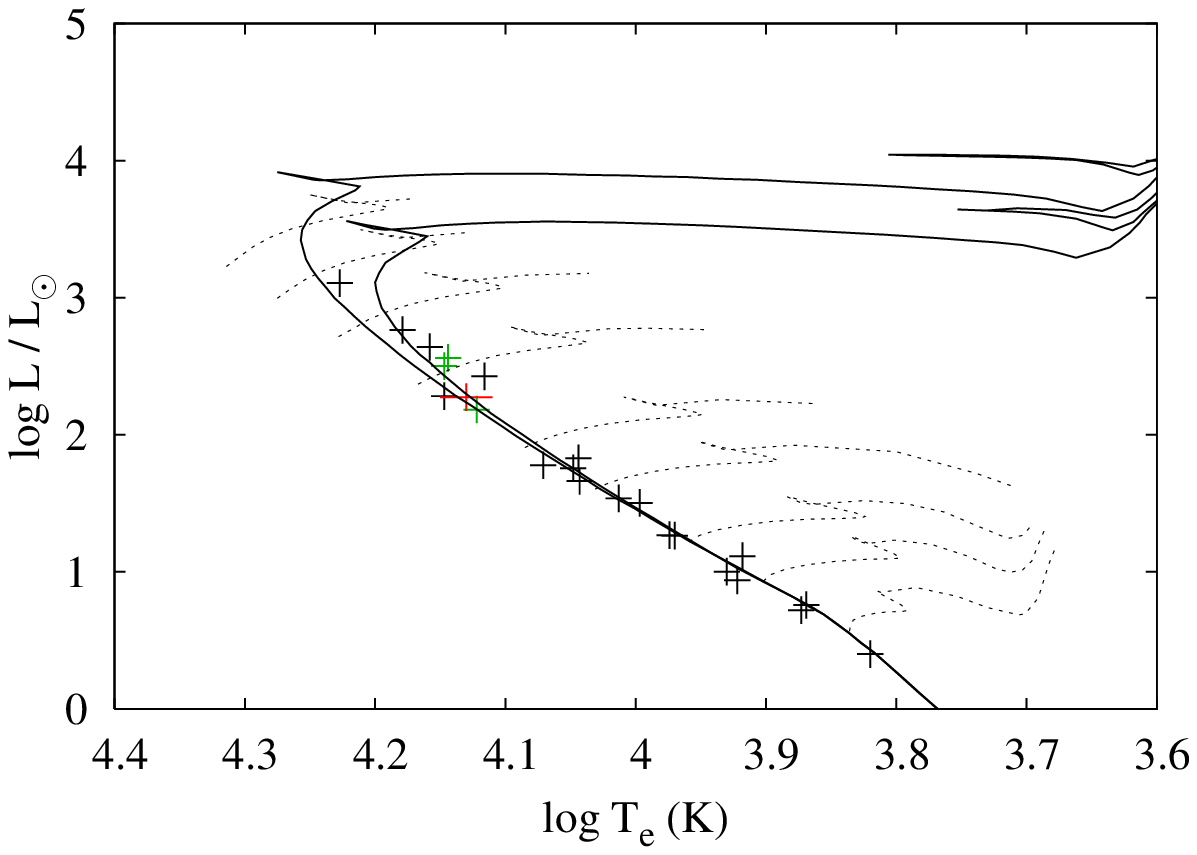}
\caption{Left Panel: Colour-magnitude diagram for NGC~2451A. Symbols
  are as in Figure~\ref{fig:ic2602-2}, with several cluster
    members that are known binaries being distinguished by filled
    green circles, and the bracketing isochrones having values of
  $\log t = 7.3$ (leftmost) and $\log t = 7.8$ (rightmost). Right
  Panel: Theoretical HRD for NGC~2451A. Symbols are as in
  Figure~\ref{fig:ic2602-2}, with the known binaries again
    distinguished by green. Bracketing isochrones corresponding to
  $\log t = 7.6$ (leftmost) and $\log t = 7.8$ (rightmost) are shown.
  Evolution tracks for $1.4, 1.7, 2.0, 2.5, 3.0, 4.0, 5.0, 6.0,$ and
  $7.0M_{\sun}$ stars are shown. }
\label{fig:2451A}
\end{figure*}

\begin{figure*}
\centering
\includegraphics*[width=3.45in]{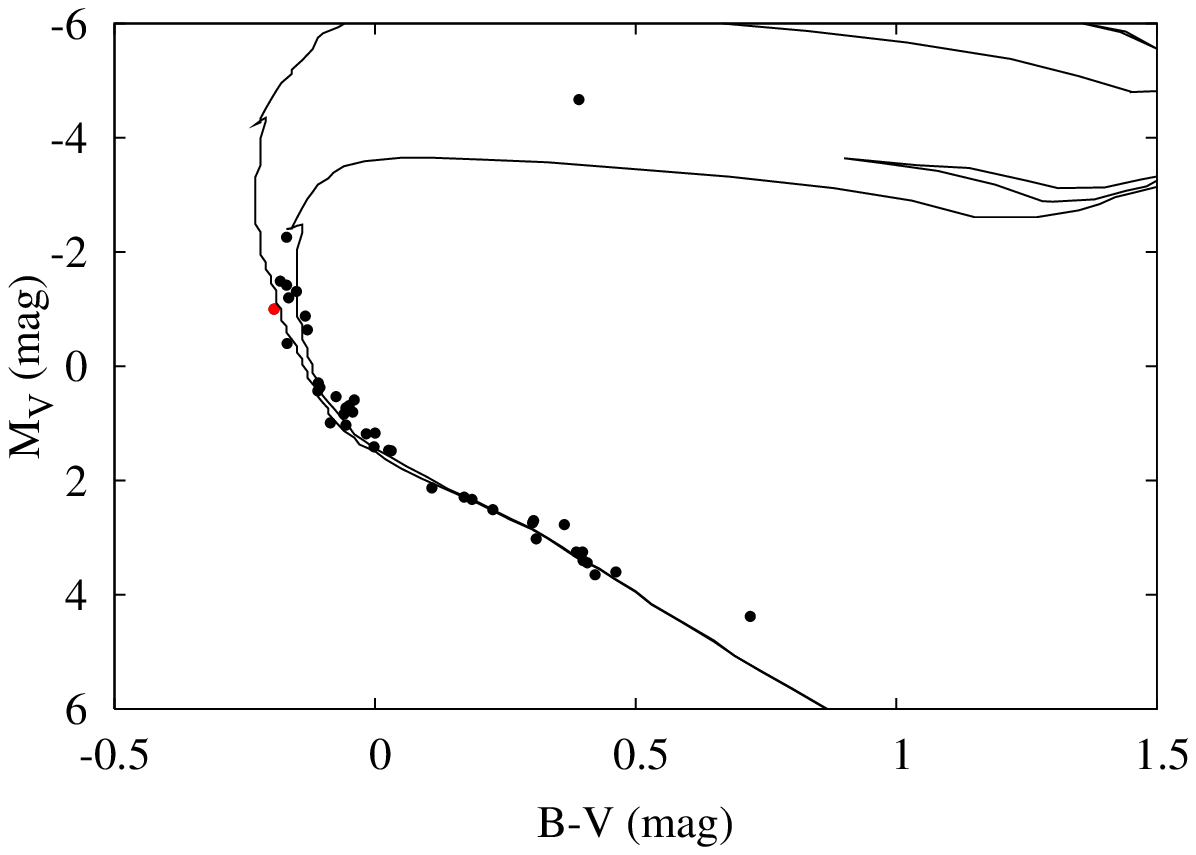}
\includegraphics*[width=3.45in]{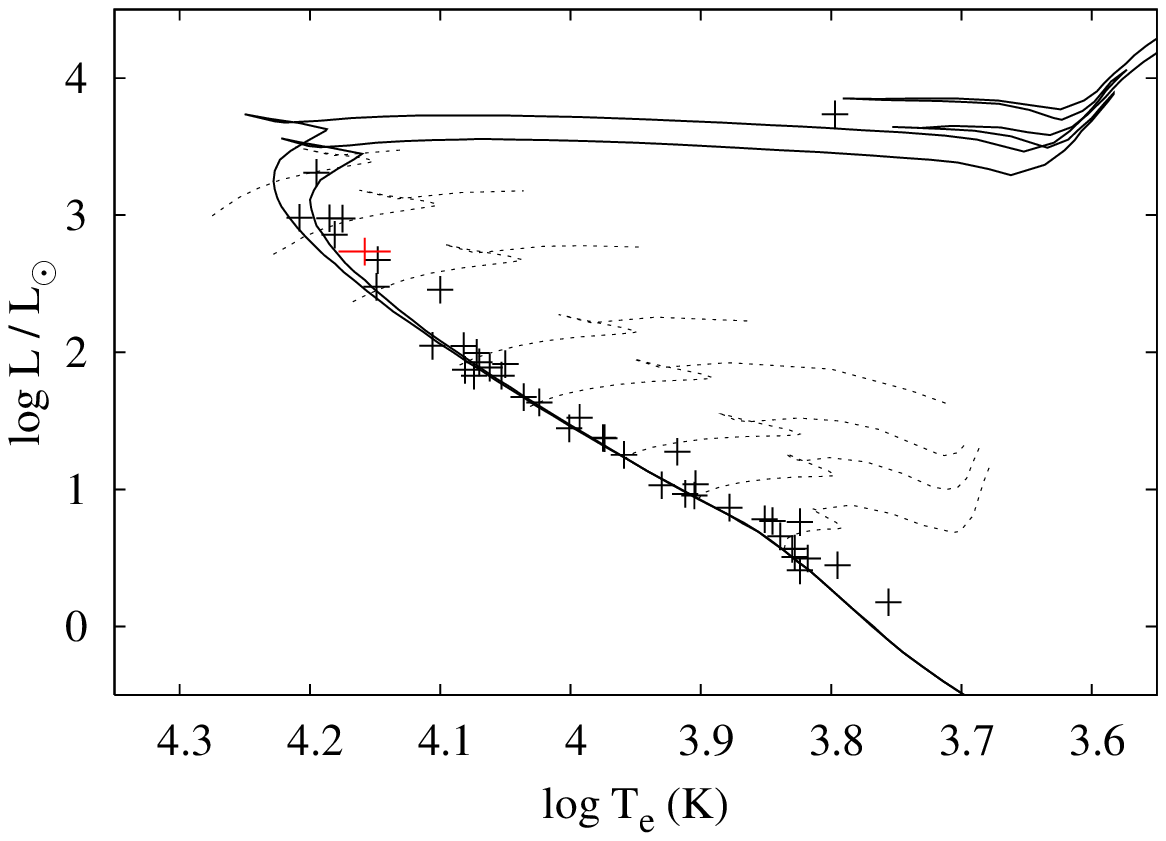}
\caption{Left Panel: Colour-magnitude diagram for $\alpha$~
  Per. Symbols are as in Figure~\ref{fig:ic2602-2} with the bracketing
  isochrones having values of $\log t = 7.3$ (leftmost) and $\log t =
  7.9$ (rightmost). Right Panel: Theoretical HRD for
  $\alpha$~Per. Symbols are as in
  Figure~\ref{fig:ic2602-2}. Bracketing isochrones corresponding to
  $\log t = 7.7$ (leftmost) and $\log t = 7.8$ (rightmost) are shown.
  Evolution tracks for $1.4, 1.7, 2.0, 2.5, 3.0, 4.0, 5.0,$ and
  $6.0M_{\sun}$ stars are shown.  }
\label{fig:aPer}
\end{figure*}

\begin{figure*}
\centering
\includegraphics*[width=3.45in]{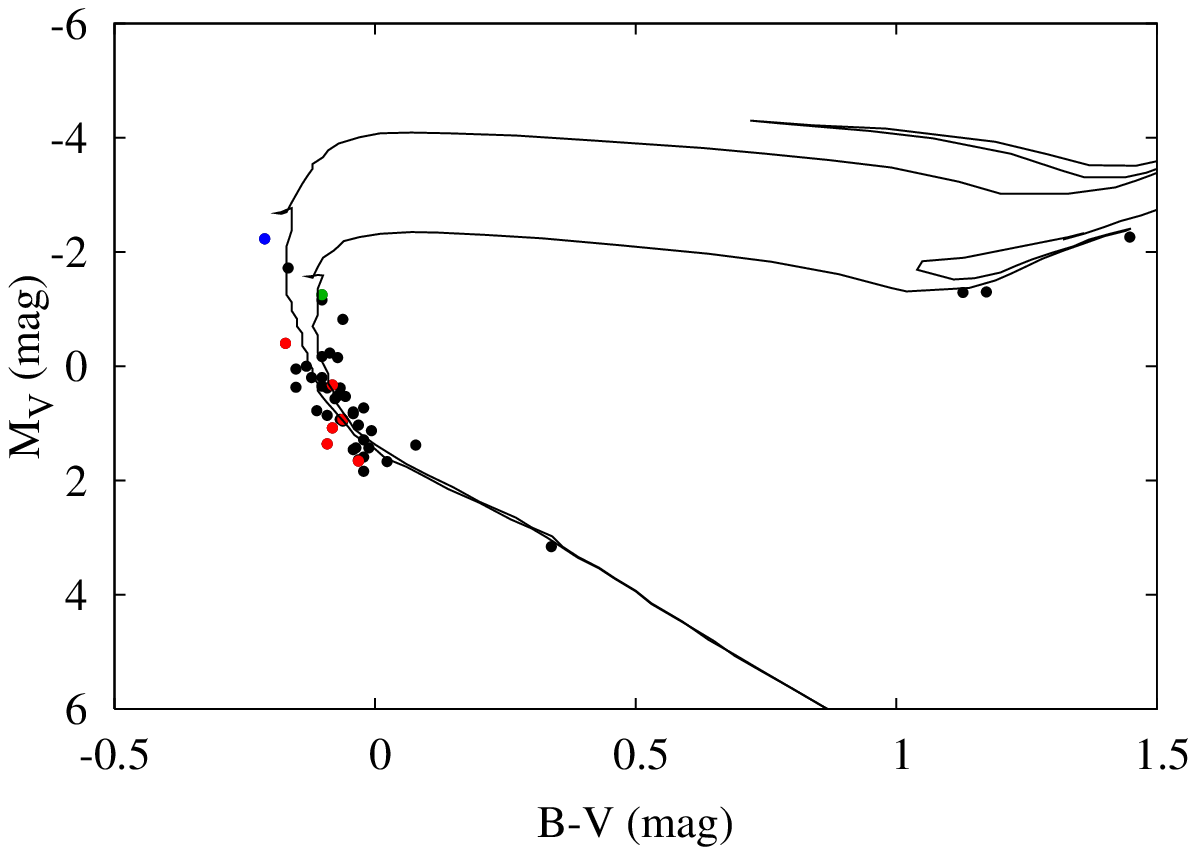}
\includegraphics*[width=3.45in]{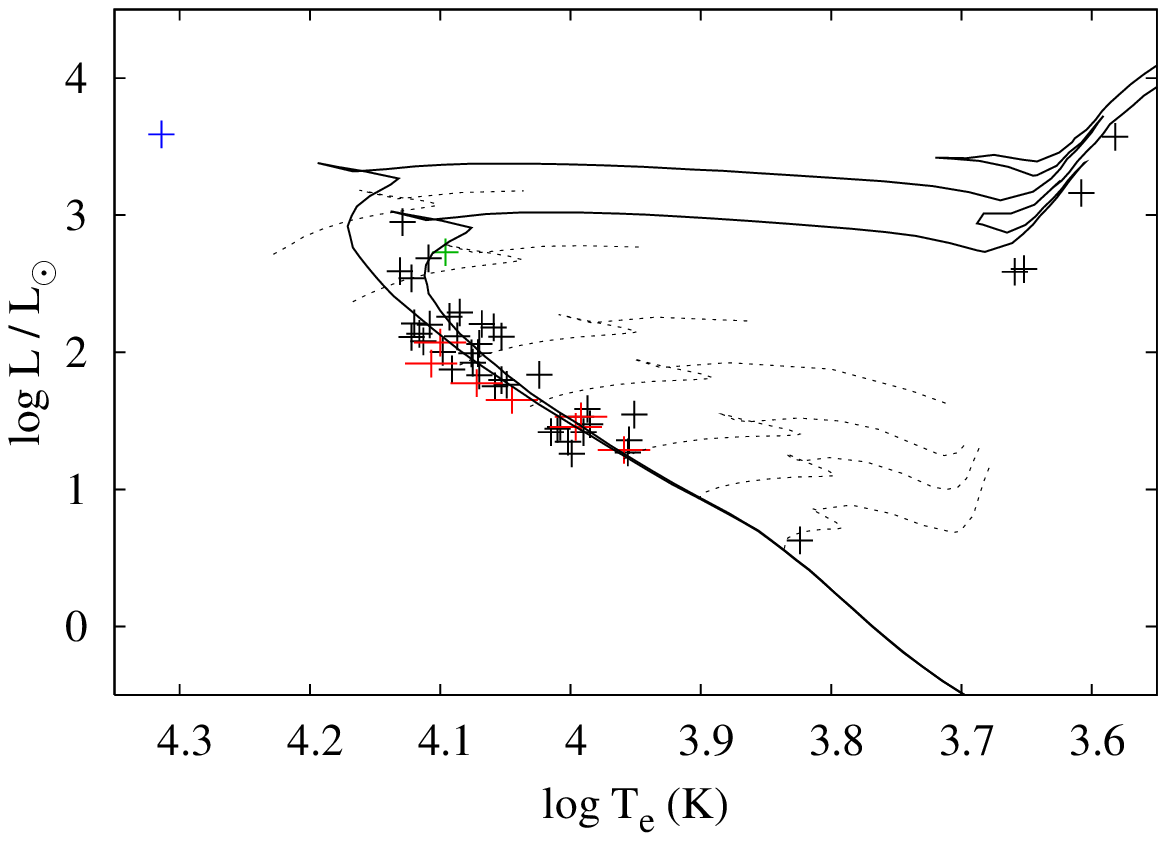}
\caption{Left Panel: Colour-magnitude diagram for NGC~2516. Symbols
  are as in Figs.~\ref{fig:ic2602-2} and \ref{fig:2451A}
  with the bracketing isochrones having values of $\log t = 7.8$
  (leftmost) and $\log t = 8.2$ (rightmost). Right Panel: Theoretical
  HRD for NGC~2516. Symbols are as in Figs.~\ref{fig:ic2602-2}
    and \ref{fig:2451A}. Bracketing isochrones corresponding
  to $\log t = 7.9$ (leftmost) and $\log t = 8.1$ (rightmost) are
  shown.  Evolution tracks for $1.4, 1.7, 2.0, 2.5, 3.0, 4.0,$ and
  $5.0M_{\sun}$ stars are shown. The apparent blue straggler is
  HD~66194.}
\label{fig:2516}
\end{figure*}

\begin{figure*}
\centering
\includegraphics*[width=3.45in]{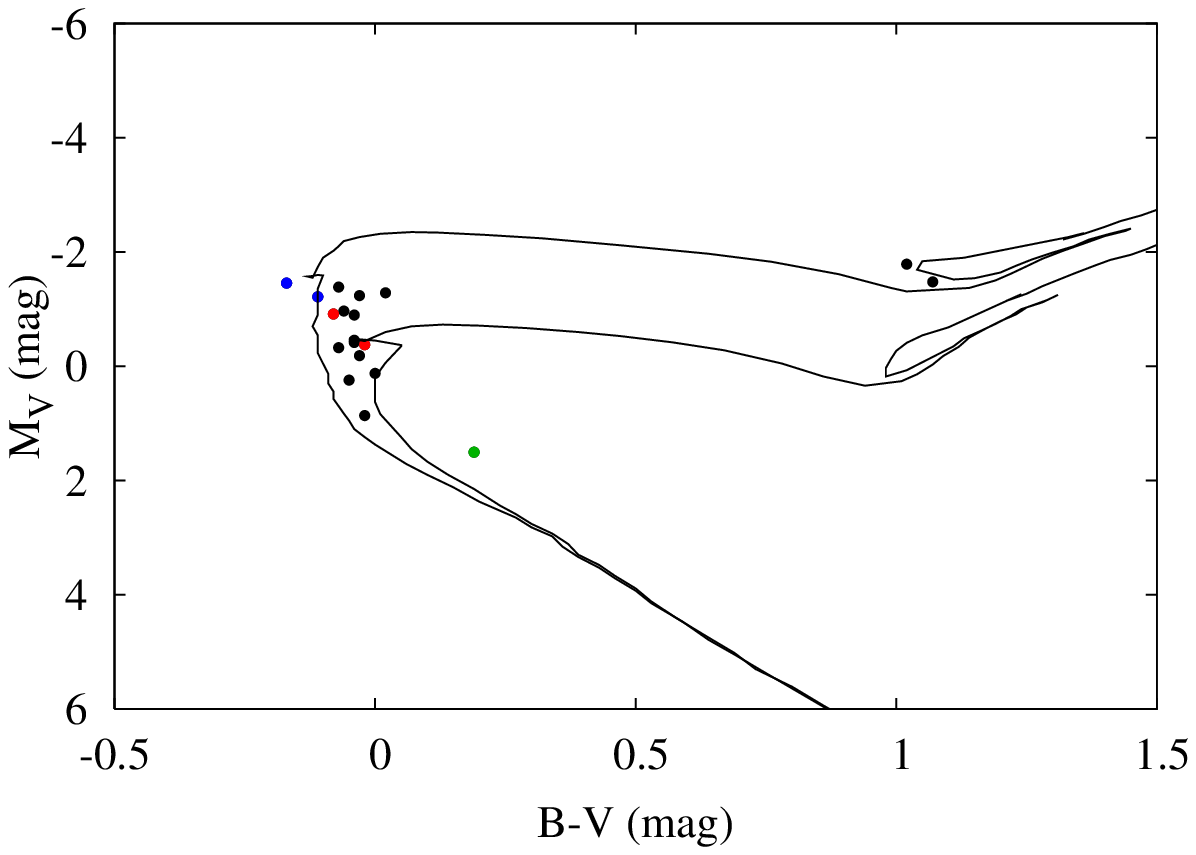}
\includegraphics*[width=3.45in]{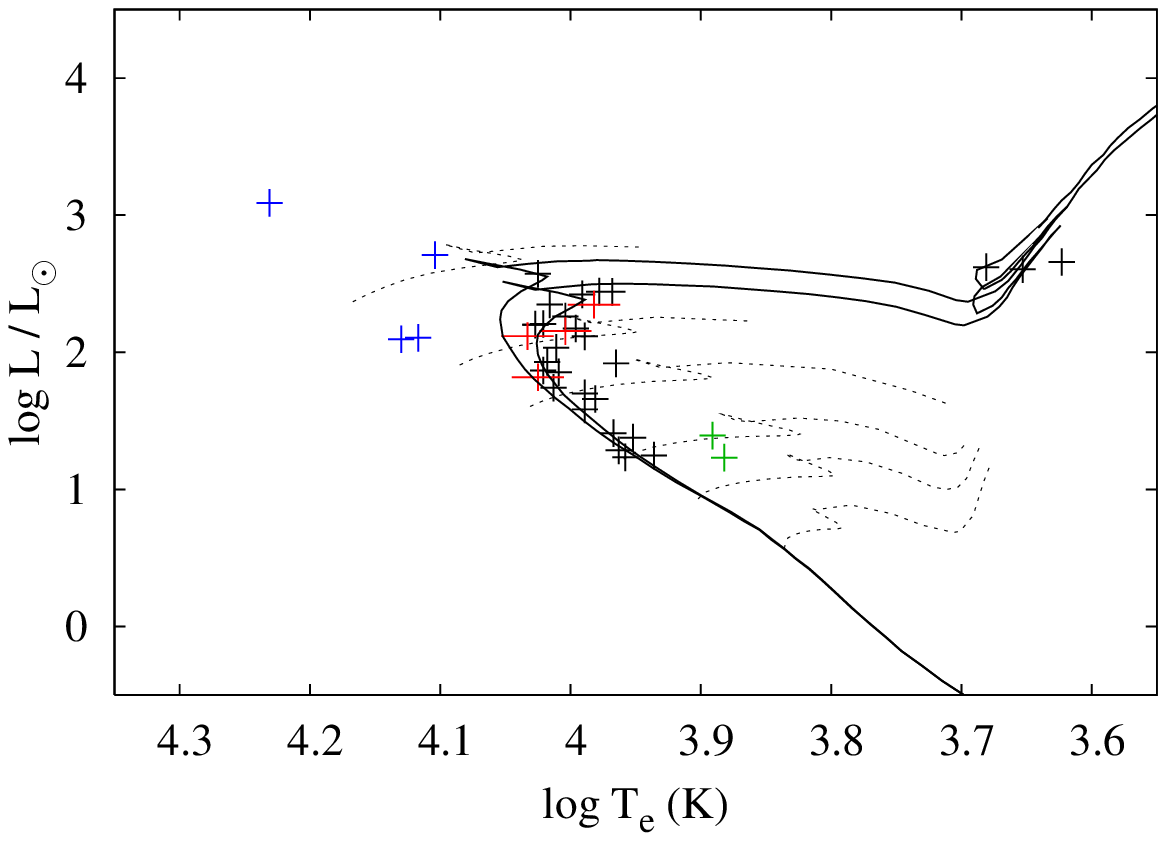}
\caption{Left Panel: Colour-magnitude diagram for NGC~6475. Symbols
  are as in Figs.~\ref{fig:ic2602-2} and \ref{fig:2451A} with the
  bracketing isochrones having values of $\log t = 8.2$ (leftmost) and
  $\log t = 8.6$ (rightmost). Right Panel: Theoretical HRD for
  NGC~6475. Symbols are as in Figs.~\ref{fig:ic2602-2} and
  \ref{fig:2451A}. Bracketing isochrones corresponding to $\log t =
  8.3$ (leftmost) and $\log t = 8.4$ (rightmost) are shown.  Evolution
  tracks for $1.4, 1.7, 2.0, 2.5, 3.0,$ and $4.0M_{\sun}$ stars are
  shown. The extreme blue straggler at $\log T_e = 4.23$ is
  HD~162374. }
\label{fig:6475}
\end{figure*}

\begin{figure*}
\centering
\includegraphics*[width=3.45in]{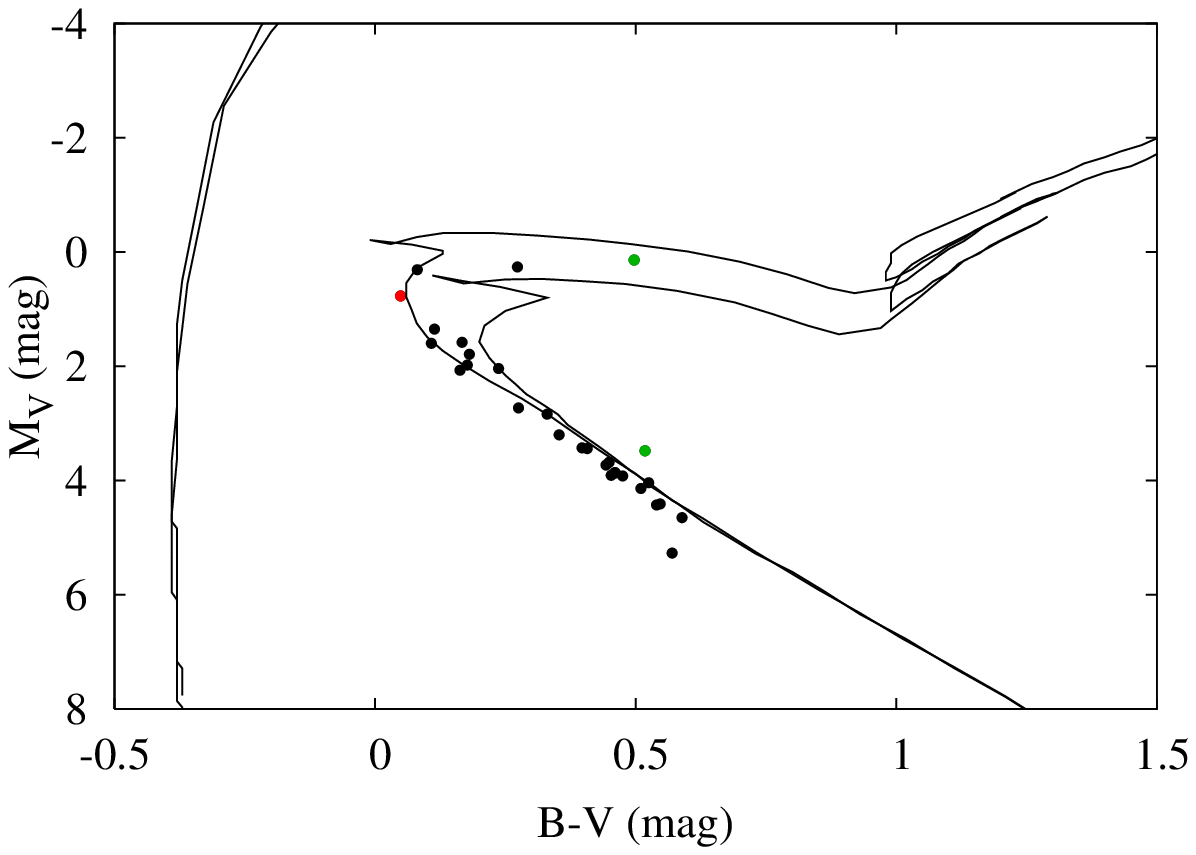}
\includegraphics[width=3.45in]{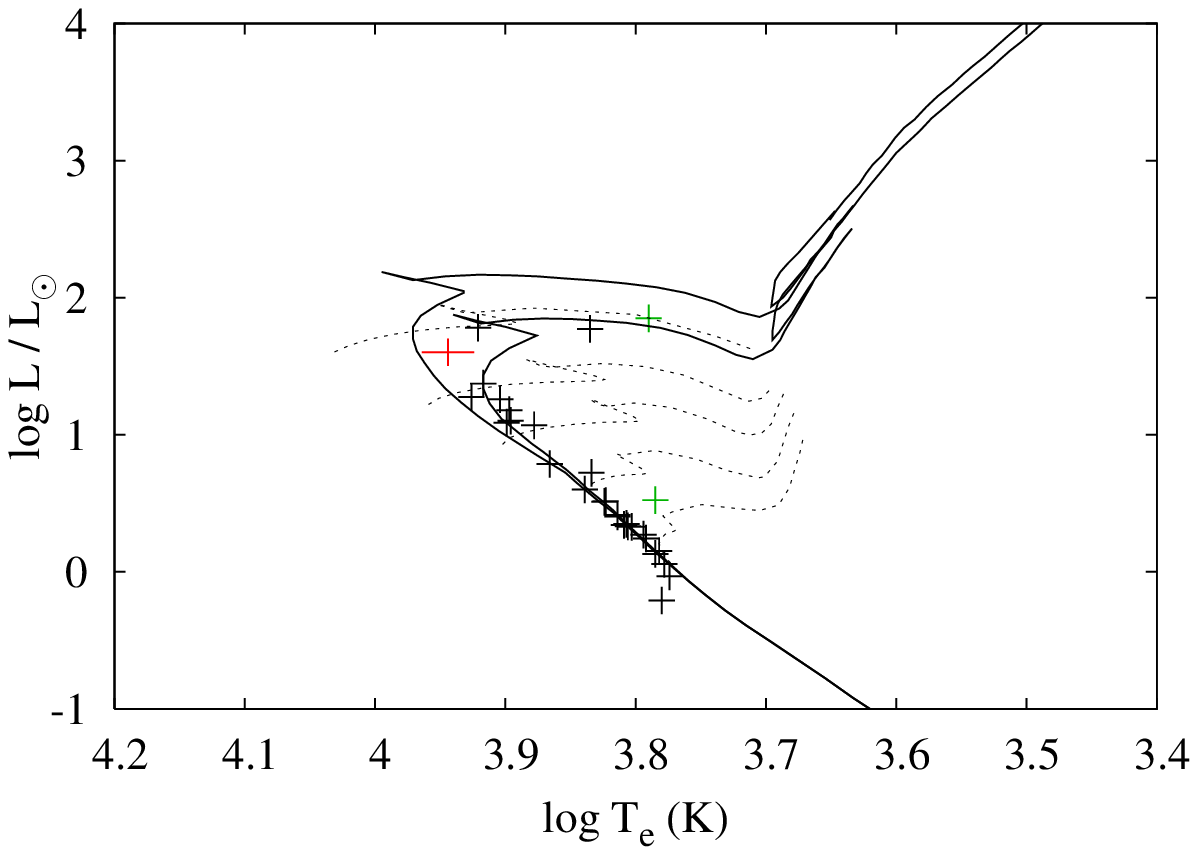}
\caption{Left Panel: Colour-magnitude diagram for Coma~Ber. Symbols
  are as in Figs.~\ref{fig:ic2602-2} and \ref{fig:2451A} with the
  bracketing isochrones having values of $\log t = 8.7$ (leftmost) and
  $\log t = 8.9$ (rightmost). Right Panel: Theoretical HRD for
  Coma~Ber. Symbols are as in Figs.~\ref{fig:ic2602-2} and
  \ref{fig:2451A}. Bracketing isochrones corresponding to $\log t =
  8.6$ (leftmost) and $\log t = 8.8$ (rightmost) are shown.  Evolution
  tracks for $1.1, 1.4, 1.7, 2.0,$ and $2.5M_{\sun}$ stars are shown.}
\label{fig:cber}
\end{figure*}

\clearpage

\begin{acknowledgements}
  We thank the two anonymous referees for very helpful comments on the paper. 
  We have made extensive use of the SIMBAD astronomical database
  (http://simbad.u-strasbg.fr/simbad/sim-fid), the WEBDA database for
  open clusters (http://www.univie.ac.at/webda/), and the University
  of Lausanne photometric catalog
  (http://obswww.unige.ch/gcpd/gcpd.html). We are pleased to acknowledge
  funding provided by the Natural Sciences and Engineering Research
  Council of Canada.
\end{acknowledgements}

\renewcommand{\thefootnote}{\alph{footnote}}

\longtab{3}{ 
\begin{longtable}{rlrrrrl}
\caption{\label{Tab_Cluster_Members} Cluster Members.}\\ 
\hline\hline
HIP & Name  & $M_V$ & $(B-V)_0$ & \te(K) & log \te (K) & log $L/L_{\sun}$ \\ 
\hline
\endfirsthead
\multicolumn{7}{c}
{\tablename\ \thetable\ -- \textit{Continued from previous page}} \\
\hline
HIP & Name  & $M_V$ & $(B-V)_0$ & \te(K) & log \te (K) & log $L/L_{\sun}$ \\ 
\hline
\endhead
\hline \multicolumn{7}{r}{\textit{Continued on next page}} \\
\endfoot
\hline
\endlastfoot \\
 Cluster & IC 2602      \\ \\
\hline 
  50102 &                 HD 88980     &  2.778 &        &  7312 & 3.864 &  0.756 \\ 
  50612 &                 HD 89903     &  1.616 &        &       &       &        \\ 
  51131 &                 HD 90731     &  1.496 &        &       &       &        \\ 
  51203 &                 HD 90837     &  2.336 &        &  8011 & 3.904 &  0.952 \\ 
  51300 &                 HD 91043     &  3.256 &        &       &       &        \\ 
  51576 &                 HD 91465     & -2.684 & -0.150 & 19291 & 4.285 &  3.717 \\ 
  52059 &        HD 92385, IC 2602 17  &  0.767 & -0.120 & 11100 & 4.045 &  1.783\tablefootmark{a} \\ 
  52116 &        HD 92467, IC 2602 18  &  1.016 & -0.010 & 10860 & 4.036 &  1.665 \\ 
  52132 &        HD 92478, IC 2602 19  &  1.601 &  0.010 &  9925 & 3.997 &  1.361 \\ 
        &        HD 92535, IC 2602 21  &  2.306 &  0.200 &  8016 & 3.904 &  0.964 \\ 
  52160 &        HD 92536, IC 2602 22  &  0.363 & -0.110 & 12114 & 4.083 &  2.024 \\ 
  52221 &        HD 92664, IC 2602 27  & -0.467 & -0.210 & 14200 & 4.152 &  2.512\tablefootmark{a} \\ 
  52261 &        HD 92715, IC 2602 28  &  0.846 & -0.060 & 10998 & 4.041 &  1.744 \\ 
  52293 &        HD 92783, IC 2602 29  &  0.751 & -0.090 & 11332 & 4.054 &  1.808 \\ 
  52328 &        HD 92837, IC 2602 31  &  1.207 & -0.040 & 10573 & 4.024 &  1.567 \\ 
  52370 &        HD 92938, IC 2602 33  & -1.169 & -0.190 & 15397 & 4.187 &  2.877 \\ 
        &        HD 92966, IC 2602 34  &  1.306 & -0.045 & 10550 & 4.023 &  1.526 \\ 
        &        HD 92989, IC 2602 35  &  1.636 &  0.000 &  9951 & 3.998 &  1.349 \\ 
 52419 & HD 93030, IC 2602 37, $\theta$ Car & -3.224 & -0.270 & 30020 & 4.477&  4.355\tablefootmark{c} \\ 
        &        HD 93098, IC 2602 38  &  1.626 &  0.010 &  9727 & 3.988 &  1.337 \\ 
  52502 &        HD 93194, IC 2602 41  & -1.176 & -0.185 & 16140 & 4.208 &  2.928 \\ 
        &        HD 93517, IC 2602 47  &  1.886 &  0.055 &  9062 & 3.957 &  1.188 \\ 
  52678 &        HD 93540, IC 2602 48  & -0.641 & -0.140 & 14209 & 4.153 &  2.583 \\ 
  52701 &        HD 93549, IC 2602 49  & -0.733 & -0.130 & 14430 & 4.159 &  2.635 \\ 
  52736 &        HD 93607, IC 2602 51  & -1.116 & -0.190 & 16882 & 4.227 &  2.951 \\ 
        &        HD 93648, IC 2602 52  &  1.896 &  0.075 &  8686 & 3.939 &  1.161 \\ 
  52815 &        HD 93738, IC 2602 54  &  0.516 & -0.020 & 11085 & 4.045 &  1.883 \\ 
  52867 &        HD 93874, IC 2602 58  &  2.234 &  0.130 &  8525 & 3.931 &  1.017 \\ 
  53016 &        HD 94174, IC 2602 63  &  1.796 &  0.060 &  9081 & 3.958 &  1.225 \\ 
  53913 &                 HD 95786     &  1.526 &  0.030 &       &       &        \\ 
  54168 &                 HD 96287     &  1.256 & -0.040 &       &       &        \\ 
\hline \\ 
 Cluster & NGC 2232  \\ \\
\hline 
  30356 &                 HD 44720     & -0.432 & -0.140 & 13960 & 4.145 & 2.481 \\ 
        &       HD 45051, NGC 2232 16  &  1.278 &  0.350 &       &       &       \\ 
        &       HD 45238, NGC 2232 12  &  0.758 & -0.060 & 11151 & 4.047 & 1.791 \\ 
  30660 &        HD 45321, NGC 2232 2  & -1.486 & -0.170 & 16432 & 4.216 & 3.071 \\ 
        &       HD 45399, NGC 2232 13  &  0.818 & -0.050 & 11526 & 4.062 & 1.796 \\ 
  30700 &        HD 45418, NGC 2232 3  & -1.152 & -0.180 &       &       &       \\ 
        &       HD 45435, NGC 2232 18  &  1.528 &  0.030 &  8856 & 3.947 & 1.318 \\ 
  30758 &        HD 45516, NGC 2232 8  &  0.158 & -0.100 & 12752 & 4.106 & 2.155 \\ 
  30761 &       HD 45532, NGC 2232 10  &  0.428 & -0.080 & 12192 & 4.086 & 2.004 \\ 
  30772 &        HD 45546, NGC 2232 1  & -2.591 & -0.200 & 18790 & 4.274 & 3.653 \\ 
  30789 &        HD 45583, NGC 2232 9  &  0.338 & -0.110 & 12700 & 4.104 & 2.079\tablefootmark{a} \\ 
        &       HD 45627, NGC 2232 15  &  1.228 &  0.040 &  8935 & 3.951 & 1.443 \\ 
  31101 &                 HD 46165     & -0.192 & -0.090 & 13610 & 4.134 & 2.359 \\ 
        &      HD 295100, NGC 2232 38  &  2.648 &  0.270 &  7292 & 3.863 & 0.807 \\ 
        &      HD 295102, NGC 2232 14  &  1.028 & -0.070 & 11168 & 4.048 & 1.684 \\ 
        &               NGC 2232 40    &  3.068 &  0.410 &       &       &       \\ 
\hline \\ 
 Cluster & NGC 2451A    \\ \\
\hline 
       &    CD-37 3859, NGC 2451A 70  &  3.677 &  0.420 &  6603 & 3.820 &  0.400 \\ 
       &    CD-38 3604, NGC 2451A 24  &  2.871 &  0.240 &  7458 & 3.873 &  0.721 \\ 
       &      HD 61621, NGC 2451A 27  &  1.740 &  0.040 &  9339 & 3.970 &  1.264 \\ 
 37297 &     HD 61831, NGC 2451 W201  & -1.509 & -0.200 & 16849 & 4.227 &  3.107 \\ 
 37322 &      HD 61878, NGC 2451A 60  & -0.641 & -0.130 & 13918 & 4.144 &  2.561\tablefootmark{b} \\ 
 37450 &      HD 62226, NGC 2451A 41  & -0.937 & -0.160 & 15115 & 4.179 &  2.765 \\ 
 37514 &      HD 62376, NGC 2451A 43  &  0.179 & -0.100 & 13252 & 4.122 &  2.184\tablefootmark{b} \\ 
       &      HD 62398, NGC 2451A 68  &  1.233 & -0.030 & 10304 & 4.013 &  1.536 \\ 
       &      HD 62479, NGC 2451A 62  &  2.775 &  0.260 &  7394 & 3.869 &  0.758 \\ 
 37557 &       HD 62503, NGC 2451A 8  &  0.909 & -0.080 &       &       &        \\ 
 37623 &                 HD 62578     & -0.751 & -0.140 & 14404 & 4.158 &  2.641 \\ 
       &      HD 62642, NGC 2451A 87  &  1.246 & -0.020 &  9921 & 3.997 &  1.503 \\ 
 37666 &      HD 62712, NGC 2451A 54  &  0.078 & -0.170 & 14032 & 4.147 &  2.282 \\ 
 37697 &      HD 62803, NGC 2451A 22  &  1.059 & -0.090 & 11050 & 4.043 &  1.663 \\ 
       &      HD 62876, NGC 2451A 67  &  2.271 &  0.110 &  8515 & 3.930 &  1.002 \\ 
 37752 &      HD 62893, NGC 2451A 78  & -0.459 & -0.130 & 13071 & 4.116 &  2.426 \\ 
       &      HD 62961, NGC2451A 15   &  1.745 &  0.030 &  9428 & 3.974 &  1.268 \\ 
       &      HD 62974, NGC 2451A 69  &  1.959 &  0.130 &  8286 & 3.918 &  1.115 \\ 
 37838 &      HD 63079, NGC 2451A 88  &  0.649 & -0.080 & 11080 & 4.044 &  1.829 \\ 
 37829 &       HD 63080, NGC 2451A 2  &  0.849 & -0.040 & 11165 & 4.048 &  1.756 \\ 
 37915 &      HD 63215, NGC 2451A 79  & -0.471 & -0.120 & 14027 & 4.147 &  2.501\tablefootmark{b} \\ 
 37982 &       HD 63401, NGC 2451A 4  &  0.001 & -0.160 & 13500 & 4.130 &  2.274\tablefootmark{a} \\ 
       &      HD 63511, NGC 2451A 38  &  2.406 &  0.150 &  8348 & 3.922 &  0.939 \\ 
 38268 &                 HD 64028     &  0.909 & -0.090 & 11765 & 4.071 &  1.778 \\ 
\hline \\ 
 Cluster & $\alpha$ Per    \\ \\
\hline 
 14697 &              Melotte 20 56   &  4.381 &  0.720 &  5708 & 3.756 &  0.176 \\ 
 14853 &   BD+49 868, Melotte 20 135  &  3.251 &  0.398 &  6723 & 3.828 &  0.568 \\ 
 15911 &   BD+46 745, Melotte 20 632  &  3.257 &  0.386 &  6754 & 3.830 &  0.565 \\ 
 15654 &   BD+47 808, Melotte 20 481  &  2.707 &  0.304 &  7100 & 3.851 &  0.782 \\ 
 15898 &   BD+47 816, Melotte 20 621  &  3.401 &  0.399 &  6734 & 3.828 &  0.507 \\ 
 15160 &   BD+48 871, Melotte 20 270  &  3.651 &  0.422 &  6671 & 3.824 &  0.409 \\ 
 15363 &   BD+49 897, Melotte 20 365  &  3.441 &  0.407 &  6582 & 3.818 &  0.495 \\ 
 16455 &   BD+49 958, Melotte 20 958  &  2.741 &  0.302 &  6996 & 3.845 &  0.769 \\ 
 16625 &  BD+49 967, Melotte 20 1050  &  3.021 &  0.309 &  6907 & 3.839 &  0.657 \\ 
 16880 &  BD+49 976, Melotte 20 1187  &  3.601 &  0.462 &  6238 & 3.795 &  0.446 \\ 
 14949 &    HD 19767, Melotte 20 151  &  2.511 &  0.226 &  7559 & 3.878 &  0.867 \\ 
 14980 &    HD 19805, Melotte 20 167  &  1.481 &  0.031 &  9443 & 3.975 &  1.375 \\ 
 15040 &    HD 19893, Melotte 20 212  &  0.691 & -0.050 & 11291 & 4.053 &  1.829 \\ 
 15259 &    HD 20191, Melotte 20 333  &  0.731 & -0.056 & 12051 & 4.081 &  1.872 \\ 
 15388 &    HD 20344, Melotte 20 379  &  1.561 &        &  8390 & 3.924 &  1.279 \\ 
 15404 &    HD 20365, Melotte 20 383  & -1.309 & -0.151 & 16143 & 4.208 &  2.982 \\ 
 15420 &    HD 20391, Melotte 20 386  &  1.481 &  0.026 &  9417 & 3.974 &  1.373 \\ 
 15444 &    HD 20418, Melotte 20 401  & -1.439 & -0.170 & 15301 & 4.185 &  2.978 \\ 
 15499 &    HD 20475, Melotte 20 421  &  2.771 &  0.363 &  6673 & 3.824 &  0.761 \\ 
 15505 &    HD 20487, Melotte 20 423  &  1.181 & -0.017 &  9842 & 3.993 &  1.523 \\ 
 15531 &    HD 20510, Melotte 20 441  &  0.591 & -0.040 & 11533 & 4.062 &  1.888 \\ 
 15556 &    HD 20537, Melotte 20 450  &  0.840 & -0.060 &       &       &        \\ 
 15770 &    HD 20809, Melotte 20 557  & -1.159 & -0.166 & 15163 & 4.181 &  2.857 \\ 
 15819 &    HD 20863, Melotte 20 581  &  0.531 & -0.075 & 11737 & 4.070 &  1.927 \\ 
 15863 &    HD 20902, Melotte 20 605  & -4.629 &  0.391 &  6270 & 3.797 &  3.736 \\ 
 15878 &    HD 20931, Melotte 20 612  &  1.411 & -0.002 & 10031 & 4.001 &  1.445 \\ 
 15988 &    HD 21071, Melotte 20 675  & -0.399 & -0.169 & 14105 & 4.149 &  2.478 \\ 
 16011 &    HD 21091, Melotte 20 692  &  1.036 & -0.056 & 10560 & 4.024 &  1.634 \\ 
 16047 &    HD 21117, Melotte 20 703  &  1.171 &  0.000 &       &       &        \\ 
 16079 &    HD 21181, Melotte 20 735  &  0.376 & -0.106 & 11804 & 4.072 &  1.995 \\ 
 16118 &    HD 21238, Melotte 20 747  &  0.466 &        & 11227 & 4.050 &  1.914 \\ 
 16147 &    HD 21278, Melotte 20 774  & -1.489 & -0.182 & 14970 & 4.175 &  2.975 \\ 
 16137 &    HD 21279, Melotte 20 775  &  0.801 & -0.043 & 11867 & 4.074 &  1.829 \\ 
 16210 &    HD 21362, Melotte 20 810  & -0.889 & -0.134 & 14073 & 4.148 &  2.672 \\ 
 16318 &    HD 21527, Melotte 20 885  &  2.331 &  0.186 &  8036 & 3.905 &  0.955 \\ 
 16340 &    HD 21551, Melotte 20 904  & -0.629 & -0.130 & 12577 & 4.100 &  2.456 \\ 
 16403 &    HD 21600, Melotte 20 921  &  2.201 &  0.109 &  8518 & 3.930 &  1.030 \\ 
 16426 &    HD 21619, Melotte 20 931  &  2.317 &  0.171 &  8163 & 3.912 &  0.966 \\ 
 16430 &    HD 21641, Melotte 20 955  &  0.301 & -0.109 & 12074 & 4.082 &  2.046 \\ 
 16470 &    HD 21699, Melotte 20 985  & -0.989 & -0.194 & 14400 & 4.158 &  2.735\tablefootmark{a} \\ 
 16574 &   HD 21855, Melotte 20 1056  &  1.731 &        &  9102 & 3.959 &  1.252 \\ 
 16782 &   HD 22136, Melotte 20 1153  &  0.431 & -0.110 & 12770 & 4.106 &  2.047 \\ 
 16826 &   HD 22192, Melotte 20 1164  & -2.209 & -0.170 & 15654 & 4.195 &  3.310 \\ 
 16966 &   HD 22401, Melotte 20 1259  &  0.996 & -0.086 & 10859 & 4.036 &  1.673 \\ 
 16995 &   HD 22440, Melotte 20 1260  &  2.121 &        &  8017 & 3.904 &  1.038 \\ 
\hline \\ 
 Cluster & NGC 2516     \\ \\
\hline 
        &    CPD-59 936, NGC 2516 117  &  1.599 & -0.022 & 10344 & 4.015 &  1.393 \\ 
        &    CPD-60 944, NGC 2516 208  &  0.339 & -0.082 & 12600 & 4.100 &  2.071\tablefootmark{a} \\ 
        &      CPD-60 945, NGC 2516 5  &  0.499 & -0.072 & 12543 & 4.098 &  2.002 \\ 
        &     CPD-60 947, NGC 2516 13  &  0.051 & -0.152 & 12814 & 4.108 &  2.202 \\ 
        &      CPD-60 948, NGC 2516 8  &  1.679 &  0.023 &  9039 & 3.956 &  1.269 \\ 
        &      CPD-60 952, NGC 2516 9  &  1.268 & -0.022 &  9660 & 3.985 &  1.475 \\ 
        &      CPD-60 961, NGC 2516 2  &  0.774 & -0.042 & 11304 & 4.053 &  1.797 \\ 
        &     CPD-60 968, NGC 2516 41  &  0.972 & -0.062 &  8935 & 3.951 &  1.545 \\ 
 120403 &     CPD-60 969, NGC 2516 11  &  0.539 & -0.057 & 10579 & 4.024 &  1.835 \\ 
        &     CPD-60 971, NGC 2516 58  &  1.655 & -0.032 & 10043 & 4.002 &  1.348 \\ 
        &      CPD-60 975, NGC 2516 1  &  0.998 & -0.032 &  9697 & 3.987 &  1.586 \\ 
        &    CPD-60 978, NGC 2516 127  &  0.925 & -0.062 & 11800 & 4.072 &  1.775\tablefootmark{a} \\ 
        &     CPD-60 979, NGC 2516 83  &  0.402 & -0.067 & 11905 & 4.076 &  1.992 \\ 
        &    CPD-60 980, NGC 2516 128  & -1.311 &  1.173 &  4490 & 3.652 &  2.607 \\ 
        &     CPD-60 985, NGC 2516 37  &  0.199 & -0.122 & 11751 & 4.070 &  2.061 \\ 
        &     CPD-60 989, NGC 2516 36  &  1.412 & -0.042 & 10188 & 4.008 &  1.456 \\ 
        &    CPD-60 990, NGC 2516 132  &  0.569 & -0.077 & 11885 & 4.075 &  1.924 \\ 
        &     CPD-60 993, NGC 2516 48  &  0.779 & -0.112 & 12337 & 4.091 &  1.875 \\ 
        &    CPD-60 1001, NGC 2516 33  &  1.453 &  0.078 &  9020 & 3.955 &  1.358 \\ 
        &    CPD-60 1029, NGC 2516 65  &  1.859 & -0.022 &  9967 & 3.999 &  1.261 \\ 
  38226 &      HD 64320, NGC 2516 225  & -1.301 &  1.128 &  4560 & 3.659 &  2.587 \\ 
  38310 &                 HD 64507     & -0.787 &        & 13526 & 4.131 &  2.591 \\ 
  38433 &                 HD 64831     & -0.175 &        & 12378 & 4.093 &  2.259 \\ 
  38536 &      HD 65094, NGC 2516 226  & -0.171 & -0.102 & 11705 & 4.068 &  2.206 \\ 
        &      HD 65467, NGC 2516 114  &  1.430 & -0.012 &  9765 & 3.990 &  1.418 \\ 
  38739 &      HD 65578, NGC 2516 116  &  0.149 & -0.102 & 12218 & 4.087 &  2.117 \\ 
  38759 &      HD 65599, NGC 2516 118  &  0.839 & -0.092 & 11206 & 4.049 &  1.763 \\ 
  38783 &      HD 65662, NGC 2516 119  & -2.281 &  1.448 &  4060 & 3.608 &  3.162 \\ 
  38779 &      HD 65663, NGC 2516 120  & -1.329 & -0.102 & 12483 & 4.096 &  2.729\tablefootmark{b} \\ 
        &        HD 65691, NGC 2516 6  &  0.909 & -0.067 & 11421 & 4.058 &  1.752 \\ 
        &      HD 65712, NGC 2516 230  &  1.359 & -0.092 &  9900 & 3.996 &  1.456\tablefootmark{a} \\ 
        &        HD 65835, NGC 2516 3  &  1.453 & -0.037 & 10226 & 4.010 &  1.442 \\ 
        &       HD 65869, NGC 2516 10  & -0.291 & -0.087 & 12169 & 4.085 &  2.290 \\ 
 120401 &       HD 65896, NGC 2516 12  &  1.159 & -0.007 &  9816 & 3.992 &  1.530 \\ 
        &       HD 65931, NGC 2516 39  &  0.769 & -0.022 & 11740 & 4.070 &  1.832 \\ 
 120402 &       HD 65949, NGC 2516 91  &  0.359 & -0.152 & 13240 & 4.122 &  2.111 \\ 
  38906 &      HD 65950, NGC 2516 126  & -1.149 & -0.102 & 12842 & 4.109 &  2.685 \\ 
 120404 &       HD 65987, NGC 2516 15  & -0.394 & -0.172 & 12600 & 4.100 &  2.364\tablefootmark{a} \\ 
        &       HD 66137, NGC 2516 19  & -0.161 & -0.072 & 11445 & 4.059 &  2.181 \\ 
  38994 &      HD 66194, NGC 2516 134  & -2.192 & -0.212 & 20632 & 4.314 &  3.589\tablefootmark{c} \\ 
        &       HD 66259, NGC 2516 20  &  0.373 & -0.092 & 11780 & 4.071 &  1.994 \\ 
        &       HD 66295, NGC 2516 26  &  1.095 & -0.082 & 11100 & 4.045 &  1.652\tablefootmark{a} \\ 
        &       HD 66318, NGC 2516 24  &  1.643 & -0.032 &  9100 & 3.959 &  1.287\tablefootmark{a} \\ 
  39073 &      HD 66341, NGC 2516 136  & -1.695 & -0.167 & 13473 & 4.129 &  2.950 \\ 
  39070 &      HD 66342, NGC 2516 110  & -2.841 &  1.638 &  3820 & 3.582 &  3.572 \\ 
        &       HD 66409, NGC 2516 23  &  0.389 & -0.102 & 12987 & 4.113 &  2.080 \\ 
  39386 &                 HD 67170     &  0.099 &        & 13175 & 4.120 &  2.210 \\ 
  39438 &                 HD 67277     &  0.272 &        & 13074 & 4.116 &  2.134 \\ 
  39879 &                 HD 68372     & -0.711 &        & 13248 & 4.122 &  2.540 \\ 
        &              NGC 2516 209    &  0.759 & -0.042 & 12800 & 4.107 &  1.918\tablefootmark{a} \\ 
        &              NGC 2516 210    & -0.021 & -0.132 & 11300 & 4.053 &  2.114 \\ 
        &              NGC 2516 227    &  3.105 &  0.338 &  6672 & 3.824 &  0.627 \\ 

  \hline \\  
 Cluster & NGC 6475     \\ \\
\hline 
  87102 &                HD 161575     & -0.423 &        & 10500 & 4.021 &  2.206 \\ 
  87134 &                HD 161649     &  0.434 &        & 13500 & 4.130 &  2.095\tablefootmark{c} \\ 
  87240 &                HD 161855     & -0.046 &  0.000 & 10250 & 4.011 &  2.033 \\ 
  87360 &     HD 162144, NGC 6475 131  &  0.275 & -0.050 & 10435 & 4.018 &  1.929 \\ 
        &       HD 162223, NGC 6475 4  &  1.604 &        &  7617 & 3.882 &  1.231\tablefootmark{b} \\ 
        &       HD 162224, NGC 6475 6  &  1.774 &        &  9083 & 3.958 &  1.234 \\ 
        &      HD 162285, NGC 6475 12  &  1.364 &        &  9264 & 3.967 &  1.410 \\ 
        &      HD 162305, NGC 6475 14  &  0.494 &        & 10600 & 4.025 &  1.817\tablefootmark{a} \\ 
        &      HD 162349, NGC 6475 24  &  1.014 &        &  9746 & 3.989 &  1.583 \\ 
  87460 &      HD 162374, NGC 6475 26  & -1.441 & -0.170 & 17004 & 4.231 &  3.089\tablefootmark{c} \\ 
  87472 &     HD 162391, NGC 6475 134  & -1.495 &  1.070 &  4800 & 3.681 &  2.621 \\ 
        &      HD 162392, NGC 6475 28  &  1.664 &        &  9177 & 3.963 &  1.284 \\ 
        &      HD 162393, NGC 6475 29  &  0.724 &        &  9761 & 3.989 &  1.700 \\ 
        &      HD 162457, NGC 6475 34  &  0.794 & -0.020 &  9581 & 3.981 &  1.659 \\ 
  87516 &                HD 162496     & -1.290 &  1.190 &  4500 & 3.653 &  2.606 \\ 
        &      HD 162514, NGC 6475 40  &  1.214 &  0.190 &  7778 & 3.891 &  1.392\tablefootmark{b} \\ 
  87529 &      HD 162515, NGC 6475 42  & -0.784 & -0.040 & 10370 & 4.016 &  2.348 \\ 
  87560 &      HD 162576, NGC 6475 55  & -0.353 & -0.020 & 10100 & 4.004 &  2.155\tablefootmark{a} \\ 
        &      HD 162586, NGC 6475 56  & -1.236 & -0.110 & 12707 & 4.104 &  2.709\tablefootmark{c} \\ 
        &      HD 162587, NGC 6475 58  & -1.606 &  1.020 &  4200 & 3.623 &  2.658 \\ 
        &      HD 162588, NGC 6475 59  & -0.126 &        & 10800 & 4.033 &  2.118\tablefootmark{a} \\ 
  87580 &                HD 162613     &  0.639 &        & 10200 & 4.009 &  1.854 \\ 
        &     HD 162678, NGC 6475 141  & -1.076 & -0.060 &  9796 & 3.991 &  2.422 \\ 
        &      HD 162679, NGC 6475 77  & -0.366 & -0.030 & 10638 & 4.027 &  2.201 \\ 
  87616 &      HD 162724, NGC 6475 86  & -1.346 & -0.070 & 10600 & 4.025 &  2.573 \\ 
  87624 &      HD 162725, NGC 6475 88  & -0.923 & -0.080 &  9600 & 3.982 &  2.347\tablefootmark{a} \\ 
  87656 &     HD 162780, NGC 6475 104  & -0.436 & -0.040 &  9902 & 3.996 &  2.174 \\ 
        &     HD 162781, NGC 6475 103  &  0.084 &  0.000 &  9221 & 3.965 &  1.919 \\ 
        &     HD 162804, NGC 6475 108  & -0.366 & -0.070 & 10648 & 4.027 &  2.202 \\ 
  87671 &     HD 162817, NGC 6475 110  & -1.236 & -0.030 &  9500 & 3.978 &  2.442 \\ 
  87686 &                HD 162874     &  0.464 &        & 10500 & 4.021 &  1.867 \\ 
  87698 &     HD 162888, NGC 6475 121  & -0.401 & -0.040 &  9750 & 3.989 &  2.117 \\ 
  87722 &                HD 162926     & -1.286 &  0.020 &  9300 & 3.968 &  2.441 \\ 
  87785 &                HD 163109     &  0.616 &        & 10300 & 4.013 &  1.741 \\ 
  87798 &                HD 163139     & -0.617 &        & 10100 & 4.004 &  2.239 \\ 
  87844 &                HD 163251     &  0.348 &        & 13095 & 4.117 &  2.105\tablefootmark{c} \\ 
  88247 &                HD 164108     &  1.444 &        &       &       &        \\ 
        &      HD 320764, NGC 6475 23  &  1.674 &        &  8635 & 3.936 &  1.247 \\ 
        &      HD 320765, NGC 6475 18  &  1.394 &        &  8945 & 3.952 &  1.377 \\ 
\hline \\ 
 Cluster & Coma Ber     \\ \\
\hline 
  59364 &   HD 105805, Melotte 111 10  &  1.310 &  0.114 &  8256 & 3.917 &  1.373 \\ 
  59399 &   HD 105863, Melotte 111 12  &  4.849 &        &  5944 & 3.774 & -0.033 \\ 
  59527 &   HD 106103, Melotte 111 19  &  3.384 &  0.397 &  6653 & 3.823 &  0.516 \\ 
  59833 &   HD 106691, Melotte 111 36  &  3.395 &  0.407 &  6668 & 3.824 &  0.511 \\ 
  59957 &   HD 106946, Melotte 111 49  &  3.162 &  0.353 &  6899 & 3.839 &  0.601 \\ 
  60025 &   HD 107067, Melotte 111 53  &  4.040 &  0.525 &  6222 & 3.794 &  0.271 \\ 
  60066 &   HD 107131, Melotte 111 60  &  1.754 &  0.181 &  7895 & 3.897 &  1.180 \\ 
  60063 &   HD 107132, Melotte 111 58  &  4.111 &  0.510 &  6195 & 3.792 &  0.244 \\ 
  60087 &   HD 107168, Melotte 111 62  &  1.567 &  0.167 &  8026 & 3.904 &  1.260 \\ 
  60123 &   HD 107276, Melotte 111 68  &  1.948 &  0.177 &  7876 & 3.896 &  1.102 \\ 
  60206 &   HD 107399, Melotte 111 76  &  4.364 &  0.547 &  6056 & 3.782 &  0.152 \\ 
  60266 &   HD 107513, Melotte 111 82  &  2.701 &  0.275 &  7340 & 3.866 &  0.787 \\ 
  60293 &   HD 107583, Melotte 111 85  &  4.609 &  0.589 &  5998 & 3.778 &  0.058 \\ 
  60304 &   HD 107611, Melotte 111 86  &  3.819 &  0.460 &  6408 & 3.807 &  0.350 \\ 
  60347 &   HD 107685, Melotte 111 90  &  3.836 &  0.461 &  6440 & 3.809 &  0.342 \\ 
  60351 &   HD 107700, Melotte 111 91  &  0.100 &  0.497 &  6169 & 3.790 &  1.850\tablefootmark{b} \\ 
  60406 &   HD 107793, Melotte 111 97  &  4.410 &  0.540 &  6096 & 3.785 &  0.131 \\ 
  60458 &  HD 107877, Melotte 111 101  &  3.673 &  0.443 &  6515 & 3.814 &  0.404 \\ 
  60490 &  HD 107935, Melotte 111 104  &  2.006 &  0.237 &  7550 & 3.878 &  1.069 \\ 
  60582 &  HD 108102, Melotte 111 111  &  3.429 &  0.518 &  6101 & 3.785 &  0.523\tablefootmark{b} \\ 
  60611 &  HD 108154, Melotte 111 114  &  3.880 &  0.453 &  6348 & 3.803 &  0.328 \\ 
  60649 &  HD 108226, Melotte 111 118  &  3.648 &  0.449 &  6511 & 3.814 &  0.415 \\ 
  60697 &  HD 108283, Melotte 111 125  &  0.235 &  0.273 &  6841 & 3.835 &  1.772 \\ 
  60746 &  HD 108382, Melotte 111 130  &  0.301 &  0.081 &  8340 & 3.921 &  1.781 \\ 
  60797 &  HD 108486, Melotte 111 139  &  1.985 &  0.163 &  7923 & 3.899 &  1.089 \\ 
  61071 &  HD 108945, Melotte 111 160  &  0.810 &  0.049 &  8800 & 3.944 &  1.602\tablefootmark{a} \\ 
  61074 &  HD 108976, Melotte 111 162  &  3.864 &  0.475 &  6405 & 3.806 &  0.332 \\ 
  61147 &  HD 109069, Melotte 111 398  &  2.860 &  0.330 &  6780 & 3.831 &  0.722 \\ 
  61295 &  HD 109307, Melotte 111 183  &  1.574 &  0.108 &  8428 & 3.926 &  1.276 \\ 
  61402 &  HD 109483, Melotte 111 192  &  5.270 &  0.570 &  6030 & 3.780 & -0.208 \\ 
\hline\hline
\footnotetext[1]{Ap star}
\footnotetext[2]{binary star}
\footnotetext[3]{blue straggler}
\end{longtable} 
}

\end{document}